\documentclass[fleqn,usenatbib]{mnras}

\usepackage{newtxtext,newtxmath}
\usepackage[T1]{fontenc}
\usepackage{ae,aecompl}

\usepackage{graphicx}	% Including figure files
\usepackage{amsmath}	% Advanced maths commands
\usepackage{amssymb}	% Extra maths symbols
\usepackage[capitalise, noabbrev]{cleveref}
\graphicspath{{figures/}}
\newcommand{\HII}{\ion{H}{ii}}
\newcommand{\EWHa}{EW(H$\alpha$)}
\crefname{section}{\S}{\SS}
\Crefname{section}{\S}{\SS}
\title[HII regions in the CALIFA survey]{HII regions in the CALIFA survey: I. Catalog presentation}

\author[C. Espinosa-Ponce et al.]{C. Espinosa-Ponce$^{1}$\thanks{E-mail: \url{mailto:cespinosa@astro.unam.mx}},
S. F. S\'anchez$^{1}$,
C. Morisset$^{2}$,
J. K. Barrera-Ballesteros$^{1}$,
\newauthor{L. Galbany$^{3}$, R. Garc\'ia-Benito$^{4}$, E. A. D. Lacerda$^{1}$, D. Mast$^{5,6}$}
\\
$^{1}$Instituto de Astronom\'ia, Universidad Nacional Aut\'onoma de M\'exico, AP 70264, 04510 Mexico City, Mexico\\
$^{2}$Instituto de Astronom\'ia, Universidad Nacional Aut\'onoma de M\'exico, AP 106, 22800 Ensenada, B. C., Mexico\\
$^{3}$Departamento de F\'isica Te\'orica y del Cosmos, Universidad de Granada, E-18071 Granada, Spain\\
$^{4}$Instituto de Astrof\'isica de Andaluc\'ia, CSIC, Apartado de correos 3004, E-18080 Granada, Spain\\
$^{5}$Universidad Nacional C\'ordoba. Observatorio Astron\'omico de C\'ordoba. C\'ordoba, Argentina.\\
$^{6}$Consejo de Investigaciones Cient\'{i}ficas y T\'ecnicas de la Rep\'ublica Argentina, Avda. Rivadavia 1917, C1033AAJ, CABA, Argentina
}

\date{Accepted XXX. Received YYY; in original form ZZZ}

\pubyear{2020}

\begin{document}
\label{firstpage}
\pagerange{\pageref{firstpage}--\pageref{lastpage}}
\maketitle

\begin{abstract}
We present a new catalog of \HII\ regions based on the integral field spectroscopy (IFS) data of the extended CALIFA and PISCO samples. The selection of \HII\ regions was based on two assumptions: a clumpy structure with high contrast of H$\alpha$ emission and an underlying stellar population comprising young stars.
The catalog provides the spectroscopic information of 26,408  individual regions corresponding to 924  galaxies,
including the flux intensities and equivalent widths of 51 emission lines covering the wavelength range between 3745-7200\AA.
To our knowledge, this is the largest catalog of spectroscopic properties of \HII\ regions.
We explore a new approach to decontaminate the emission lines from diffuse ionized gas contribution. This diffuse gas correction was estimated to correct every emission line within the considered spectral range. With the catalog of \HII\ regions corrected, new demarcation lines are proposed for the classical diagnostic diagrams.
Finally, we study the properties of the underlying stellar populations of the \HII\ regions. It was found that there is a direct relationship between the ionization conditions on the nebulae and the properties of stellar populations besides the physicals condition on the ionized regions.
\end{abstract}

\begin{keywords}
ISM: HII regions -- ISM: general -- galaxies: ISM -- galaxies: star formation -- techniques: spectroscopic
\end{keywords}

%%%%%%%%%%%%%%%%%%%%%%%%%%%%%%%%%%%%%%%%%%%%%%%%%%

%%%%%%%%%%%%%%%%% BODY OF PAPER %%%%%%%%%%%%%%%%%%

\section{Introduction}
Classical \HII\ regions are clouds of ionized gas surrounding young massive stars in which the star formation has recently taken place ($<15$Myr). The ultraviolet photons that ionize the surrounding medium are produced in large amounts by OB stars that were formed in the \HII\ regions. Since OB stars only live $\sim 10$ Myr, the \HII\ regions trace recent star formation processes. These regions possess a broad range of physical size (diameter), from a few parsecs, such as the Orion nebula with $D\lesssim 8$ pc \citep{ander14}, to hundreds of parsecs, like 30 Doradus with $D\sim20$ pc, NGC 604 with $D\sim460$ pc, or NGC 5471 with $D\sim1$ kpc \citep{oey03, rgb2011}. The smaller regions are ionized by a single star or a handful of them, while the largest ones are ionized by a young cluster of thousand of them. Those larger star-forming complex are the precursors of the extragalactic giant \HII\ regions found commonly in the disks of spiral galaxies \citep[e.g.,][]{hodg83, dott87,dott89, knap98}, or in starbursts and blue compact galaxies \cite[][]{kehrig08,angel09,cairos12}.

In addition to \HII\ regions, there are other emission-line objects with different main excitation mechanisms: planetary nebulae and objects photoionized by hard radiation fields such as active galactic nucleus (AGN), shock excitation or even post-AGB stars \citep[e.g., ][]{veilleux87, binn94, 2003Stasinska_A&A397, binn09, mori09, 2011Flores-Fajardo_MNRAS415, papa13, sign13}. All these objects, including \HII\ regions, could be classified, in principle, based on their emission line ratios. The first diagnostic diagram using the emission lines ratios [NII]$\lambda6583/$H$\alpha$ vs. [OIII]$\lambda5007/$H$\beta$, was proposed by \citet[][hereafter BPT diagram]{baldwin81}. \cite{osterbrock89} and \cite{veilleux87} proposed a demarcation curve between AGN and star-forming galaxies and two new diagnostic diagrams\footnote{The two more commonly-used diagnostic diagrams are [SII]$\lambda 6717,6761$/H$\alpha$ vs. [OIII]$\lambda 5007$/H$\beta$ and [OI]$\lambda 6300$/H$\alpha$ vs. [OIII]$\lambda 5007$/H$\beta$.}. Theoretical photoionization models have been used to infer the demarcation line between the different ionization sources.

\cite{dopita00} and \cite{kewley01} proposed one of the first classification schemes to identify pure galaxies hosting star-formation and AGNs, using a combination of stellar population synthesis and photoionization models. Their models assume that the emission line spectrum depends directly on: (i) the chemical abundances of heavy elements (oxygen being the most important) of the gas phase within an \HII\ region; (ii) the shape of ionizing radiation spectrum; (iii) the electron density; and (iv) the geometrical distribution of gas with respect to the ionizing sources. All geometrical factors are included in the so-called ionization parameter $q$ (defined as the ratio of ionizing photon density to hydrogen density) or its dimensionless counterpart, $U$\footnote{The ionization parameter is defined as $U = S/(4\pi r^2 n_H c)$, where $S$ is the number of ionization photons emitted by a source per unit of time, $r$ is the distance between the source and the nebula, $n_H$ is the H density and $c$ is the speed of light}. They assume {\it a priori} that these parameters are independent one each other and, hence, the models are presented as grids of the four of them. On the other hand, \cite{kauffmann03} use the Sloan Digital Sky Survey \citep[SDSS, ][]{york00} data to propose a different demarcation line to separate ionization by star-forming (SF) and AGNs, based purely on observational properties.

\HII\ regions allows the present chemical composition of the interstellar medium (ISM) to be determined.. They have been used for a long time to measure the gas-phase abundance directly at discrete locations in galaxies \citep[e.g.,][]{diaz89, 2013Marino_aa559A, rosales11, laura16}. These measurements provide the necessary information to understand the distribution of chemical abundances across the optical extension of galaxies, to constrain different theories of galactic chemical evolution, to obtain information about the stellar nucleosynthesis in star-forming galaxies and to derive the star-formation histories of galaxies.

Classical studies of the chemical abundance of extragalactic nebulae are based on the observations of nebular emissions through long-slit or single-aperture spectrographs. 
Unfortunately, these techniques provided a limited sample of galaxies in detail, including a low number of \HII\ regions per galaxy, in general. Therefore, there was a limited coverage of these regions along the whole galaxies extension \citep[e.g.,][]{bresolin19}. Conversely, the classical slit-spectroscopy surveys introduce a bias by aperture effects: long-slit observations do not integrate over the full extension of an \HII\ region. Instead, only the brightest region of each nebulae are observed, i.e., the areas of higher excitation. Hence, the derived spectra may be not representative of the whole ionization conditions \citep[e.g.,][]{dopita14}.

%Older 
Moreover, traditional surveys could present a biased sample of \HII\ regions. Most of these spectroscopic surveys are biased toward high-contrast (thus, equivalent width of emission lines) nebulae located in the outer disk of galaxies. Since late-type galaxies offer the best contrast with their lower surface-brightness in the continuum, this type of galaxies were mainly observed by these surveys. Thus, the final samples and conclusions could be biased to the properties of \HII\ regions in a particular type of galaxy. Moreover, in most of the cases, the studies were restricted to regions ionized by younger and/or massive stellar clusters.
In general, the central regions of galaxies were excluded, although it is known that they distinguish themselves spectroscopically from those in the disk: inner star-forming regions present stronger low-ionization forbidden emission and may populate the right branch of the BPT diagram \citep{kennicutt89, ho97,sanchez12b}.

Despite these caveats, these studies provide important relations, clear patterns and scaling laws e.g., characteristics vs integrated abundances: \citealt{moustakas06}; effective yield vs. luminosity and circular velocity relations \citep[e.g.,][]{Garnett:2002p339}; abundance gradients and the effective radius of disks \citep[e.g.,][]{diaz89}; possible differences in the gas-phase abundance gradients between normal and barred spirals \citep{zaritsky94,Martin:1994p1602}; surface brightness vs metallicity relations, luminosity-metallicity and mass-metallicity \citep{leque79,skill89, VilaCostas:1992p322, zaritsky94, tremonti04}, to cite just a few.

In the last years, new observational techniques, as Integral Field Spectroscopy (IFS) with large field-of-view (FoV) and multi-object spectroscopy, allow us to observe hundreds of \HII\ regions per galaxy and a full two-dimensional (2D) coverage of the disk of nearby ones \citep[e.g.,][]{rosales-ortega10, sanchez12a, sanchez14}. This new generation of emission-lines surveys provided large catalogs with thousands of objects over an unbiased sample of galaxies, from early to late type galaxies \citep{2016A&A...585A..47M, laura16}. Moreover, the sampled \HII\ regions are at any galactocentric distances (even at more than 2 $R_e$)\footnote{The effective radius, $R_e$ of a galaxy is defined as the radius at which half of the total light of the galaxy is emitted} which allows to study a wider range of \HII\ regions.

The Calar Alto Legacy Integral Field Area survey \citep[CALIFA, ][]{sanchez12a} was one of the pioneering in the massive exploration of \HII\ regions over complete samples of galaxies \citep[e.g.,][]{sanchez14}. CALIFA provides IFS data for a well defined and  statistically significant sample of galaxies on the nearby Universe ($z<0.08$). The observations cover the full optical extent (up to $~3-4R_e$) of the galaxies of any morphological type and with a spectral range $3745-7200$\AA. CALIFA still provides the best physical resolution \citep[$\sim$0.8 kpc,][]{2019arXiv191106925S} compared to other IFS galaxy surveys \citep[e.g.,][]{sami, manga} due to its limited redshift range and lower average redshift. However, it is overpassed by IFS observations with MUSE (e.g., L\'opez-Coba et al. submitted), nonetheless, this instrument does not cover wavelengths bluer than 4650 \AA.
In \cite{sanchez12a} the first \HII\ regions catalog derived from the observations of the pilot study of the CALIFA survey was presented \citep[p-CALIFA;][]{marmol-queralto11}. Besides that, several studies were performed with a larger statistical sample of \HII\ regions based on later catalog updates: oxygen abundance profiles on face-on spiral galaxies \citep{laura16, 2016Marino_A&A585A}, global and local M-Z relations and the dependence on the star formation rate (SFR) and the specific star formation rate (sSFR) \citep{sanchez13, 2017Sanchez_MNRAS469}, etc. However, there was no public update of the CALIFA \HII\ regions catalog after \cite{sanchez12a}.

In this study, we present the updated CALIFA catalog of \HII\ regions for $\sim 1000$ nearby galaxies observed so far within the framework of this survey (August 2019). We propose new criteria to select \HII\ regions within each galaxy and a new methodology to extract their main spectroscopic information. We explore the distribution of \HII\ regions along the classic diagnostic diagrams. For the first time we do not use those diagrams to classify \HII\ regions. Instead,  we present an unbiased exploration of the location of these objects across the diagnostic diagrams. Moreover, we propose a new approach to correct the contamination by diffuse ionized gas (DIG). With the final catalog of \HII\ regions corrected by diffuse gas, we explore how this correction changes the distribution along those diagnostic diagrams. Finally we show how \HII\ regions distribute among galaxies of different morphologies and at different galactocentric distances, revising patterns with the properties of the underlying stellar population already covered in previous explorations. In addition we deliver the full catalog of \HII\ regions\footnote{\url{http://ifs.astroscu.unam.mx/CALIFA/HII\_regions/}}, including all the spectroscopic properties extracted by our code. The code itself is also delivered for its use by the community\footnote{\url{https://github.com/cespinosa/pyHIIexplorerV2}}. 

The layout of this article is as follows: in Section \cref{sec:sample} we present the galaxy sample and data analyzed in this study; the performed analysis is described in \cref{sec:ana}, including a summary of the procedure adopted to extract the information of the emission of stellar population and the emission lines (\cref{sec:Pipe3D}), and a detailed description of the procedure to detect and extract that information for the \HII\ regions within each galaxy (\cref{sec:det}). The final criteria adopted to select the \HII\ regions are described in \cref{sec:sel}, and the proposed procedure to de-contaminate by the DIG emission are described in \cref{sec:diff}. The main results of our analysis are presented in \cref{sec:results}, including: (i) the main effects of the DIG decontamination (\cref{sec:DIG_effect}); (ii) the new proposed demarcation lines to distinguish \HII\ regions from other sources of ionization (\cref{sec:demarc}); (iii) the distribution of \HII\ regions by morphology (\cref{sec:morph}) and (iv) by galactocentric distance (\cref{sec:dist}); and (v) the revisited relations between the properties of the underlying stellar population and the line ratios in \HII\ regions (\cref{sec:age_met}). Finally, we show the demarcations line for the other two most used diagnostic diagrams in \cref{sec:app_demar_lines}. We describe how to access the catalog and the code in the \cref{sec:cat}. And, we show the equivalent width of H$\alpha$, \EWHa,  for the diffuse ionized gas sample in \cref{app:DIG}.

\section{Sample and data}
\label{sec:sample}

We selected the analyzed galaxies from the extended CALIFA (eCALIFA); the pilot CALIFA (pCALIFA) and PISCO samples \citep{sanchez12a, DR3, 2018Galbany_ApJ855}. eCALIFA comprises all galaxies observed by the CALIFA survey \citep{sanchez12a}, and all the extensions to the original sample observed using the same configuration and almost similar selection criteria \citep[as described in][]{DR3}. pCALIFA comprises the galaxies observed by the CALIFA survey in its pilot phase. PMAS/PPak Integral-field Supernova hosts COmpilation \citep[PISCO,][]{2018Galbany_ApJ855} sample comprises IFS data of 232 supernova host galaxies. The three surveys (eCALIFA, pCALIFA and PISCO) were observed with the 3.5 m telescope at the Calar Alto Observatory (CAHA).
eCALIFA $+$ pCALIFA $+$ PISCO comprises an homogeneous dataset in terms of setup, depth and redshift range. It includes galaxies selected in a similar way (basically matching the FoV of the instrument) observed along the last 9 years without a fixed observing schedule. We based the current results on the 924 galaxies observed up to August 2019. We will call then eCALIFA dataset for simplicity along this article hereafter.

The details of CALIFA survey, sample selection, observational strategy, and reduction are explained in \cite{sanchez12a}. The details of PISCO sample are explained in \cite{2018Galbany_ApJ855}. The whole galaxy sample were observed using the PMAS spectrograph \citep{roth05} in the PPAK configuration \citep{kelz06}. The instrument FoV has a hexagonal shape of $74'' \times 64''$. In order to map the full optical extent of the galaxies up to $\sim$2.5R$_e$ within this FoV, the galaxies of mother sample were selected by their diameter \citep{walcher14}, a primarily criteria adopted for all extended sub-samples. Details of the particular selection criteria for each CALIFA (\& PISCO) extension are given in \citet{DR3}. The complete coverage of the FoV is guaranteed  with the observing strategy described in \cite{DR3}, with a final spatial resolution of $FWHM \sim 2.4''$, corresponding to $\sim $0.8 kpc at the average redshift of the survey. The spectroscopic resolution and sampled wavelength range for the V500 setup (the currently adopted in this study) are 3745-7200\AA\ and $\lambda/\Delta\lambda \sim 850$, respectively. This resolution and range are sufficient to analyze the most important ionized gas emission lines from [OII]$\lambda3727$ to [SII]$\lambda6731$ in the redshift range of our galaxy sample and to deblend and subtract the underlying stellar population (e.g., \citealt{kehrig12, sanchez12a, cid-fernandes13,cid-fernandes14,Pipe3D_I}). 

The data were reduced using version 2.2 of the {CALIFA} pipeline. The modifications with respect to the early versions presented in \cite{sanchez12a}, \cite{husemann13}, and \cite{2015GarciaBenito_A&A576A} are described in detail in \cite{DR3}. In summary, the data fulfill the predicted quality-control requirements with a spectrophotometric accuracy that is better than 5\% everywhere within the explored wavelength range, both absolute and relative, with a depth that allows us to detect emission lines in individual \HII\ regions as faint as $\sim 10^{-17}$ erg s$^{-1}$ cm$^{-2}$ and with a signal-to-noise of S/N$\sim 3-5$. For the strong emission lines considered in the current study, the S/N is well above this limit , and the measurement errors are negligible in most of the cases \citep[as described in][]{2015Sanchez_AA573A}. Unavoidable, S/N of the weak emission lines, e.g., [OIII]$\lambda4363$, could be less than $\sim 3-5$. In all cases, they have been propagated and included in the final error budget.

The final product of the data reduction is a regular-grid datacube, with two spatial dimensions ($x$ and $y$ coordinates corresponding to the right ascension and declination of the target) and one spectral dimension ($z$ coordinate corresponding to a common step in wavelength). The CALIFA pipeline also provides a proper mask cube of bad pixels, the propagated error cube and a prescription of how to handle the errors when performing spatial binning (due to covariance between adjacent pixels after image reconstruction). We use as the starting point of our analysis these datacubes, together with the ancillary data described in \cite{walcher14}.

\section{Analysis}
\label{sec:ana}

\subsection{Stellar population and emission line properties}
\label{sec:Pipe3D}

Spatially resolved stellar population and emission line properties provided by the {\sc Pipe3D} pipeline \citep[developed in purpose for the analysis of IFS data, ][]{Pipe3D_II} are used in this work. This pipeline has been extensively used in the study of CALIFA \citep[e.g.,][]{laura16,carlos19,jairo19}, MaNGA \citep[e.g.,][]{ibarra16,sanchez18, 2016Barrera_MNRAS463, 2017Barrera_ApJ844, 2018Barrera_ApJ852}, SAMI \citep[e.g.,][]{sanchez19} and MUSE \citep[][]{carlos17} datasets. {\sc Pipe3D} adopts {\sc FIT3D } as the core fitting package \citep{Pipe3D_I}, and the GSD156 simple stellar population library in the current implementation. This library that includes 156 templates: 39 stellar ages (from 1 Myr to 14.1 Gyr), and four metallicities (Z/Z$\odot$=0.2, 0.4, 1, and 1.5), has been already used in many different previous publications \citep[e.g.,][]{perez13,rosa14,ibarra16,jairo19}.

Details of the fitting procedure, the adopted dust attenuation curve, and limitations of the processing of the stellar populations are given in \citet{Pipe3D_I,Pipe3D_II}. We provide here a summary for understanding the nature of the properties explored in the current study. First, a spatial binning is performed in each datacube to increase the S/N of the continuum (S/N$\sim$50 per \AA\ at 5000\AA) at any location within the FoV. This limit was selected as a compromise between not losing significant spatial information and providing accurate properties of the stellar populations. It is based on the simulations presented in \citet{Pipe3D_I}. As shown in \citet{sanchez12a} and \cite{DR3},
a significant fraction of the original spaxels ($\sim$50\%) have an S/N above this goal, those remain to unbind. For the remaining ones, the adjacent spaxels are aggregated, co-adding the corresponding spectra, to increase the S/N. However, contrary to other binning schemes \citep[e.g., Voronoi binning, ][]{capp03}, the adopted one only aggregates those spaxels which relative flux intensities at the selected wavelength range (5000\AA) differ less than 15\% from each other. In this way, the original shape of the galaxy is better preserved, and there is no mixing of nearby regions that correspond to different structures (e.g., arm/inter-arms). The penalty is that the goal S/N is not reached for all the final bins, that in general, have a S/N well above 40 for the CALIFA data \citep[see][for some examples of the procedure]{ibarra16}. In average, $\sim$1000-2000 tessella (binned regions) are obtained and their corresponding spectra for each CALIFA datacube.

After the binning process, the stellar population is fitted for each binned spectra following two steps. As the first step, the stellar velocity and velocity dispersion, and the stellar dust attenuation, are derived using a limited sub-set of the stellar library described before. As the second step, a multi-SSP linear fitting is performed, using the full GSD156 library, adopting the kinematics and dust attenuation derived in the previous step. A Monte-Carlo iteration is performed for the second step, perturbing the original spectra with their corresponding errors, to provide errors on the estimated stellar parameters. This procedure gives a stellar-population model for each tessella. Then, we derive a stellar-population model for each spaxel by re-scaling the model within each bin to the flux intensity of the continuum at the corresponding spaxel  \citep[within the considered bin,][]{cid-fernandes13,Pipe3D_I}. Using this model, we estimated the stellar population properties in each spaxel, generating maps of each property. The most relevant parameter considered in this article is the percentage of light and mass (weights) corresponding to each stellar population template within the library. In particular the fraction of young and old stellar populations (separated by a specific age limit, as we discuss later). Besides, several properties of the stellar component such as the stellar mass density, light-weighted (LW) and mass-weighted (MW) average stellar age (Age$_{{\rm LW,MW}}$) and metallicity ($[{\rm Z/H}]_{{\rm LW,MW}}$), and the average dust attenuation (together with the kinematic properties already described before) are derived. The maps of each of these properties are packaged as channels (slices) of two datacubes (for convenience), comprising the average stellar properties (SSP cubes) and the weights of the decomposition of the stellar population (SFH cubes), as described in \citet{Pipe3D_II}.

%%%%%%%%%%%%%%%%%%%%%%%%%%%%%%%%%%%%%%%%%%%%%%%%%%%%%%%%%%%%%%%%%%%%%%%%%%%%%%%%%%%%%%%%%%%%%%%%%%%5
\begin{figure*}
	\includegraphics[width=\textwidth]{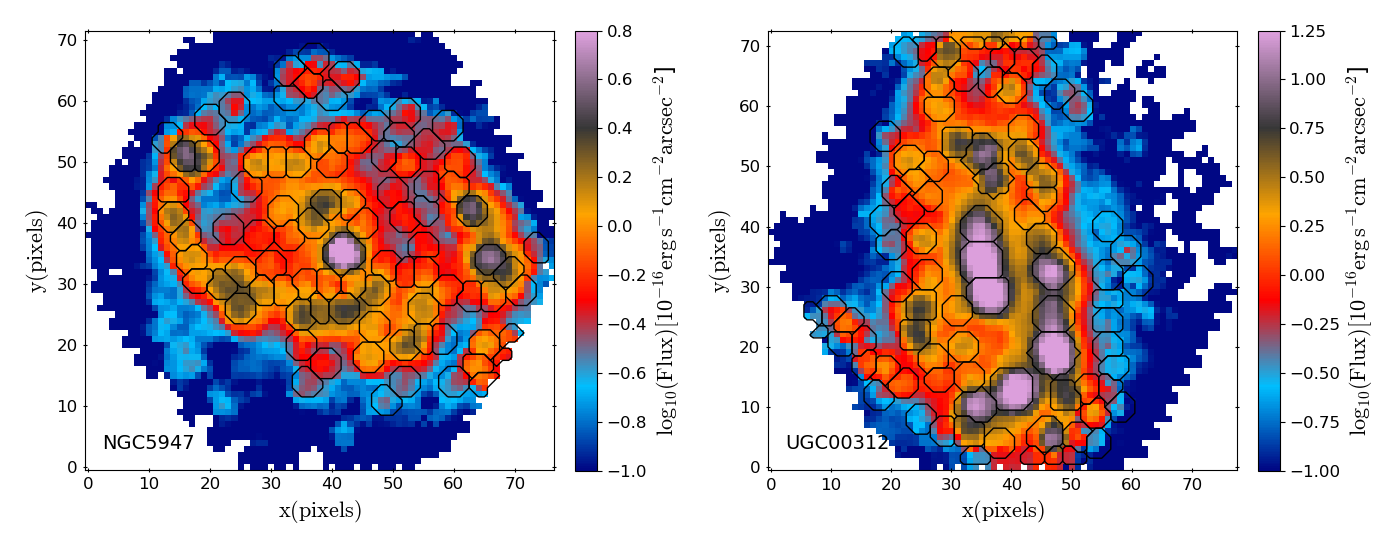}
    \caption{H$\alpha$ emission maps of NGC\,5947 and UGC\,00312 of the analyzed sample, the color code represents flux intensity. Black contours show the ionized regions found by \textsc{pyHIIexplorer} for each galaxy. }
    \label{fig:seg_map_contours}
\end{figure*}
%%%%%%%%%%%%%%%%%%%%%%%%%%%%%%%%%%%%%%%%%%%%%%%%%%%%%%%%%%%%%%%%%%%%%%%%%%%%%%%%%%%%%%%%%%%%%%%%%%%%%

In order to derive the properties of the ionized gas emission lines, for each spaxel, the stellar spectrum model (as described above) is subtracted from the original one creating a ``gas-pure cube''. This cube contains the emission lines together with the noise and the residuals of the stellar population modeling. The main properties of a set of 51 emission lines is then derived for each spaxel based on a weighted momentum analysis \citep[described in detail in][]{Pipe3D_II}, including for each line the integrated flux intensity, the velocity, velocity dispersion, and the equivalent width (EWs). The set of emission lines comprises the most relevant ones detected in the optical spectrum of \HII\ regions (see a complete list in \citealt{sanchez07c}), as detected in the Orion nebulae using a similar wavelength range. The final dataproducts of this analysis, described in \citet{Pipe3D_II} and \citet{sanchez18}, comprise a datacube for each galaxy in which each channel (slice) contain the spatial distribution (map) of each of the derived properties for each analyzed emission line (plus the corresponding maps for their errors). These dataproducts (labeled as \texttt{flux\_elines} cubes), together with the ones storing the properties of the stellar populations, are the primary input for the further analyses performed in this article\footnote{A subset of these products were distributed for a subsample of the current analyzed galaxies in \citet{Pipe3D_II}}.

\subsection{HII detection and extraction}
\label{sec:det}

We performed the segregation of \HII\ regions using \textsc{pyHIIexplorer}\footnote{\url{https://github.com/cespinosa/pyHIIexplorerV2}}. This code is based on \textsc{HIIexplorer}, originally written in \textsc{Perl} \citep{sanchez12b}. \textsc{pyHIIexplorer} does essentially the same procedures of \textsc{HIIexplorer}, re-coding it in \textsc{python}, which is more commonly used nowadays in astronomy. This makes a more easy to distribute, install, (semi-) automatic update and the integration with other codes or packages. We optimized this new version to use all the advantages of data processing that \textsc{python} offers. Finally, we implement pyHIIexplorer to run in parallel so that it can process several galaxies at the same time, depending on the available number of cores.

The detection of ionized regions performed by \textsc{pyHIIexplorer} is based on two assumptions. 1) \HII\ regions have strong emission lines that are clearly above the continuum emission and the average ionized gas emission across each galaxy. 2) the typical size of \HII\ regions is about a few hundreds of parsecs, which corresponds to a usual projected size of a few arcsec at the distance of our galaxies. These assumptions will define clumpy structures with a high H$\alpha$ emission line contrast in comparison to the continuum.

Thus, the main input of the algorithm is an H$\alpha$ emission map (preferentially, as it is the most intense emission line in \HII\ regions in the optical regime). The process to identify ionized regions consists of two steps: i) the identification of the emission peaks (i.e., local maximum) from the input emission map, and ii) the aggregation of adjacent pixels to each peak in order to segment the image. For this, the algorithm requires some additional input parameters: (i) a minimum flux intensity threshold for the peak emission; (ii) a minimum relative fraction with respect to the peak emission to keep aggregating nearby pixels; (iii) a maximum distance to the ionized region peak emission to stop the aggregation; (iv) a minimum absolute intensity threshold in the adjacent pixels to continue the aggregation.

%%%%%%%%%%%%%%%%%%%%%%%%%%%%%%%%%%%%%%%%%%%%%%%%%%%%%%%%%%%%%%%%%%%%%%%%%%%%%%%%%%%%%%%%%%%%%%%%%%%%%%%%%%%%%%%%%%%%%%%%%%%
% Fraction of young vs. EW
\begin{figure*}
	\includegraphics[width=\textwidth]{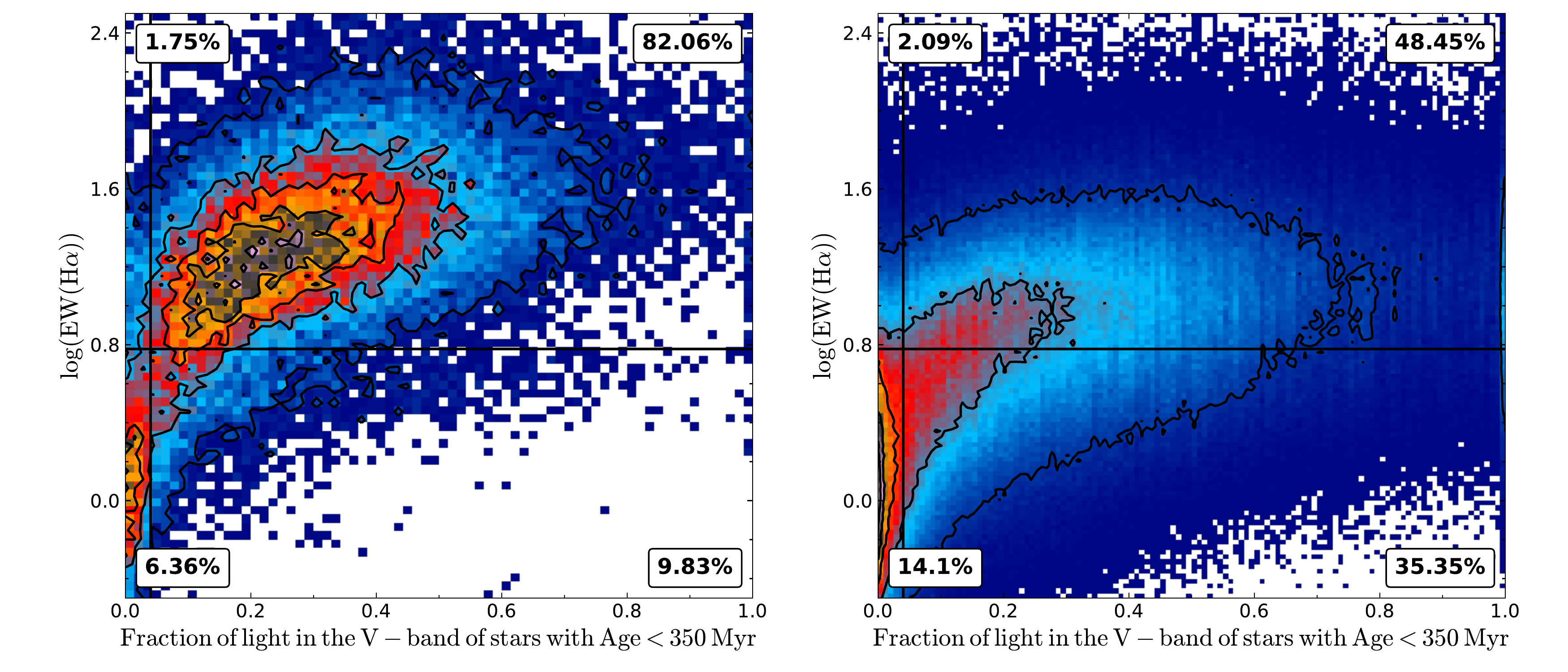}
    \caption{\EWHa\ vs the fraction of light in the \textit{V-band} with age lower than $350$Myr for the underlying stellar population for the 38,807 detected ionized regions (\textit{left panel}) and for the pixels excluded by the segregation code (\textit{right panel}), i.e., the diffuse (not clumpy) regions. The light fractions are based on the SSP analysis performed by \textsc{Pipe3D} \citep{Pipe3D_II}. The vertical and horizontal lines (in both panels) represent the adopted limits for the \EWHa\ and fraction of young stars to identify \HII\ regions. The contours (and color code) show the density distribution of regions (or spaxels). The outermost contour encloses 95\% of the regions and each consecutive one encloses the 75\%, 55\% and 35\% of the points. Finally, in both panels we show the percentage of number of ionized regions (\textit{left panel}) and DIG spaxels (\textit{right panel}) for each subdivision according our limits in \EWHa\ and fraction of light.}
    \label{fig:EW_frac_star}
\end{figure*}

%%%%%%%%%%%%%%%%%%%%%%%%%%%%%%%%%%%%%%%%%%%%%%%%%%%%%%%%%%%%%%%%%%%%%%%%%%%%%%%%%%%%%%%%%%%%%%%%%%%%%%%%%%%%%%%%%%%%%%%%%%%

 The first, second and fourth parameters are intensity lower limits, and the third one is a physical-geometry parameter. The peak (local maximum) emission must be higher than the first parameter to define a new \HII\ region. 
 The algorithm starts looking for a peak (local maximum) and aggregates the adjacent pixels following the described criteria until no further one fulfills them. Then it looks for a new peak emission and iterates the process until no new local maximum is found. Finally, it iterates again over the segregation maps to redistribute pixels to the nearest peak when two or more adjacent \HII\ regions overlaps following the previous criteria.

All the parameters described above can be derived by visual inspection of the  H$\alpha$ emission map or with a statistical analysis of the same emission maps  (i.e., determining the 3$\sigma$ intensity threshold, the typical H$\alpha$ flux of an \HII\ region to our galaxy distance, and their typical size or the PSF resolution). In this particular implementation, for the \textsc{python} version, we adopted the same set of parameters introduced for the \textsc{Perl} one \citep{sanchez12a}. It allows us to use the same numerical values for parameters in both versions and make a direct comparison between the results.

The output of the algorithm is a FITS file that contains a segmentation map (see \cref{fig:seg_map_contours}) similar in all ways to the one provided by other object detection algorithms like \textsc{SExtractor} \citep{1996Bertin_AA117}. Each ionized region is identified by a unique ID value starting with 1. The pixels that are not associated with any ionized regions have a zero as ID. These areas are regions without emission, or their emission does not fulfill the criteria described before.
Besides, a second FITS file is generated containing a masked map where all \HII\ regions identified previously have been masked. 
The structure of both FITS files are described in \cref{sec:cat}.

Finally, we extract the flux values of every emission line for each ionized region. For this, we calculate an H$\alpha$ luminosity-weighted mean with the fluxes in the original cubes (i.e., the dataproducts of \textsc{Pipe3D} described in \cref{sec:Pipe3D}) for all spaxels with the same ID-index in the derived segmentation map. The final flux values and their associated errors are stored in tables for each ionized region ordered by rows (see \cref{sec:cat}). Also, we extract the properties and the weights of the decomposition of the underlying stellar population (see \cref{sec:Pipe3D}) for each ionized region and store it in two other tables with similar structures as the flux tables (see \cref{sec:cat}).

For this particular study, the input parameters used to identify the ionized regions from the H$\alpha$ emission maps of all galaxies analyzed here are: i) a flux intensity threshold for each ionized regions peak emission of $3 \times 10^{-17} \mathrm{erg}\;\mathrm{s}^{-1} \mathrm{cm}^{-2}\mathrm{arcsec}^{-1}$, ii) a minimum relative flux to the peak emission for associated spaxels corresponding to the same \HII\ regions of 5\%, iii) a maximum distance to the location of the peak of $5.5''$ and iv) an absolute flux intensity threshold of $0.5 \times 10^{-17} \mathrm{erg}\;\mathrm{s}^{-1} \mathrm{cm}^{-2}\mathrm{arcsec}^{-1}$ in the adjacent pixels to associate them to the peak emission. All these parameters were selected by a try-and-error procedure on a handful of galaxies, in order to maximize the number of detected \HII\ without including clear diffuse ionized regions, based on the experience of previous studies \citep[e.g.,][]{sanchez12b,2015Sanchez_AA573A,laura16,laura18}. These parameters were derived by perform a visual inspection and a statistical analysis of the input emission line map, in order to maximize the number of detected ionized regions for the bulk sample of galaxies.

The current catalog was generated by running \textsc{pyHIIexplorer v2} on a computer system with an AMD\textregistered\ Ryzen\textregistered\ 2700x CPU with a $3.7$ Ghz clock speed. This CPU has 8 cores (and 16 threads) that can directly address 16GB of RAM. The complete generation of the catalog running \textsc{pyHIIexplorer} in parallel mode took $\sim25$ minutes.

\subsection{HII regions catalog}
\label{sec:sel}

We apply the procedure outlined in the previous section over 924  datacubes corresponding to a similar number of galaxies observed within the eCALIFA + pCALIFA and PISCO survey. A total of 38,807 clumpy ionized regions were identified. In order of segregate the \HII\ regions, from the clumpy ionized regions sample, we select those that fulfill the following criteria: (i) a minimum S/N$>3$ in the H$\alpha$ flux; (ii) a H$\alpha$/H$\beta$ ratio above $2.7$, to reject those regions with un-physical Balmer ratios; and finally (iii) a minimum EW(H$\alpha$)$= 6$\AA\ and (iv) a fraction of young stars to the total luminosity ($f_y$) above of the 4\% was applied, following \citet{sanchez14}.

The cuts in the \EWHa\ and $f_y$ were selected based on the previous knowledge of the properties of \HII\ regions and after a detailed inspection of the distribution of both parameters \citep[e.g.,][]{2018Lacerda_MNRAS474, 2015Sanchez_AA573A, mori16}. \cref{fig:EW_frac_star} shows the distribution of \EWHa\ against the fraction of light in the V-band of stars younger than $350$ Myr ($f_y$) for (\textit{left panel}) all clumpy ionized regions selected by the code and (\textit{right panel}) all spaxels that does not belonged to any clumpy ionized region (i.e., not selected by the code). The vertical and horizontal lines correspond to the percentage of young stars equals 4\%, and the demarcation limits of \EWHa$ = 6$\AA, respectively. These limits were selected as a compromise between preserving the largest number of \HII\ regions and removing the largest number of ionized regions that are not compatible with being an \HII\ region (i.e., lack of enough young stars). Indeed, the actual limits correspond roughly to the location of the contour including 95\%\ of the clumpy regions with young stars (upper-right) and with a lack of young stars (bottom-left). 
Those regions identified as \HII\ regions (upper-right zone in the figure) represents the 67.7\% of the total clumpy ionized regions detected by the code.\footnote{In \cref{sec:ratios_check}, we compare the distribution of [OIII] $\lambda 5007 / \lambda4959$ and the [NII] $\lambda 6584 / \lambda6548$ flux emission lines ratios for the \HII\ regions with the theoretical values.}

Applying all the criteria described above, we recover 26,408  \HII\ regions. Unlike other selection methods to obtain \HII\ regions commonly used in the literature, our method is based only on well-know properties of \HII\ regions: a clumpy ionized gas structure with high H$\alpha$ emission and a young underlying stellar population. The first criterion is satisfied by construction, it is a fundamental part of the detection algorithm described in \cref{sec:det}. The second one is provided by the adopted \EWHa\ and $f_y$ cuts that ensure that the ionized gas is indeed dominated by a young stellar population ($f_y$) and has a high contrast in the H$\alpha$ emission (\EWHa). 

\subsection{The Diffuse Ionised Gas}
\label{sec:diff}

It is well known that galaxies exhibit a low-intensity emission-line spectrum that it is broadly distributed along all their optical extension, known as diffuse ionized gas (DIG). Different authors have studied the DIG in the literature \citep[e.g.,][]{1991Reynolds_IAUS144,  1996Rand_ApJ462, 1998Rand_ApJ501, 1997Minter_ApJ484, 1997Minter_ApJ485, 2017Zhang_MNRAS466, 2018Lacerda_MNRAS474}. Unfortunately, its physical origin and properties are still not fully understood. It is detected far from the galactic plane as well as close to it \citep[e.g.,][Levy in prep.]{2011Flores-Fajardo_MNRAS415}, and in galaxies of any morphological type \citep[e.g.,][]{papa13, sign13}. The first detection of DIG was in the areas adjacent to classical \HII\ regions on the Galactic disc \citep{1971Reynolds_PhDT}; however, later works proved the existence of DIG at large distances above the galaxy planes \citep[e.g.,][]{1996Hoopes_AJ112, 1999Hoopes_ApJ522}. \cite{2007Oey_ApJ661} showed that diffuse ionized gas emission is present in galaxies of all types, and it may represent up to 60\% of the total H$\alpha$ flux intensity \citep{2002Relano_ApJ564}. However, other authors reduce its contribution to $\sim$4-10\%\, when flux intensity is corrected by the dust attenuation \citep[i.e., its contribution to the total luminosity;][]{sanchez12b}.

In the literature, the ionizing source of the DIG has been widely studied by different authors, too \citep[e.g.,][]{1996Ferguson_AJ111, 1998Rand_ApJ501}. The radiation from massive OB stars leaked out from \HII\ regions is one of the most widely accepted candidates \citep{2009Haffner_RvMP81}. Another one is ionization by hot low-mass evolved stars \citep[HOLMES][]{2011Flores-Fajardo_MNRAS415, 2012Stasinska_IAUS283} or post-AGBs stars \citep{papa13}. Finally, low-velocity shocks could contribute somehow to the DIG ionization \citep{2006Monreal-Ibero_ApJ637,2010Monreal-Ibero_AA517A,1996Dopita_ApJS102}, although its contribution is still not well estimated. Most probably, the ionizing source is different depending on the galaxy type, and on the location within galaxies \citep[][]{2018Lacerda_MNRAS474, 2019arXiv191106925S}. In more early-type galaxies and in the retired areas of late-type ones, the two later ionizing sources are maybe the most common ones \citep[e.g.,][]{sign13}. On the other hand, in more late-type galaxies and in the vicinity of star-forming areas, the first proposed one is more likely to happens \citep{2000Zurita_AA363,2002Zurita_AA386,2002Relano_ApJ564}.

As indicated before, DIG may be present anywhere in galaxies, and in particular, juxtaposed and mixed with the ionization of \HII\ regions. This may contaminate the emission lines detected at that location, and alter the line ratios if the ionization is different than the one produced by nearby young massive stars themselves. The effects of DIG in low-resolution IFS data were explored by \citet{mast14} and \citet{zhang16}, demonstrating that it can critically modify physical derivations like the oxygen abundance \citep[recently addressed by][]{asari19}. The physical spatial resolution of the eCALIFA dataset is, on average better than the resolution at which the spatial contamination effect its critical. However, these results highlight the need to attempt a correction of the DIG contamination in the emission line intensities derived for the detected \HII\ regions.

Unfortunately, it is not possible to determine and decouple the contribution of the DIG in the \HII\ regions from the individual spectra themselves: (i) the emission by the \HII\ regions are in general stronger, and it opaques the emission of the DIG; (ii) at the spectral resolution of the data, both emissions have similar kinematic properties, precluding any decoupling. A tentative approach would be to estimate this contribution using the DIG emission at the adjacent areas of the detected \HII\ regions and interpolate it. However, there are several potential problems in this approach. First, if the DIG is ionized by photons leaked from the \HII\ regions themselves, then removing its contribution is unnecessary (at a first-order), since it should show very similar emission properties, just re-scaled to lower intensity values. Second, due to the limited spatial resolution of our data, the wings of the PSF may create a considerable contribution to the adjacent areas. Again, this ionization corresponds to the \HII\ regions and should not be removed. In order to decontaminate the DIG and preserve the contribution due to young-stars, we characterize properties and determine the dominant ionizing source of DIG.

\subsubsection{Characterizing the DIG ionizing source}
\label{sec:DIG_char}

A commonly adopted procedure to segregate the DIG ionized by old-stars and
the star-forming (SF) regions is using the H$\alpha$ surface brightness ($\Sigma_{{\rm H}\alpha}$) since it is directly related to the density of ionized gas \citep{2007Oey_ApJ661, 2017Zhang_MNRAS466}. However, this criterion is valid only for face-on thin galaxies. Due to projection effects and for galaxies with any thickness $\Sigma_{{\rm H}\alpha}$ becomes an extensive quantity, since the light is co-added along the line-of-sight for the same observed area. This was recently discussed by \citet{2018Lacerda_MNRAS474}, who proposed a classification scheme based on the equivalent width of H$\alpha$. This approach relies on the results by \citet{cid-fernandes10} and \citet{sanchez14}, and on the fact that the \EWHa\ is always an intensive quantity. According to them the ionized gas could be classified in three groups: (i) areas with \EWHa$<3$\AA\ correspond to diffuse gas being ionized predominantly by an old stellar population (HOLMES/post-AGBs); (ii) areas with \EWHa$>14$\AA\ correspond to the diffuse gas that its ionized by OB stars; and (iii) $3$\AA$<$\EWHa$<14$\AA\, correspond to the diffuse gas that is powered by a mixture of both ionization sources.

%%%%%%%%%%%%%%%%%%%%%%%%%%%%%%%%%%%%%%%%%%%%%%%%%%%%%%%%%%%%%%%%%%%%%%%%%%%%%%%%%%%%%%%%%%%%%%%%%%
\begin{figure}
	\includegraphics[width=\columnwidth]{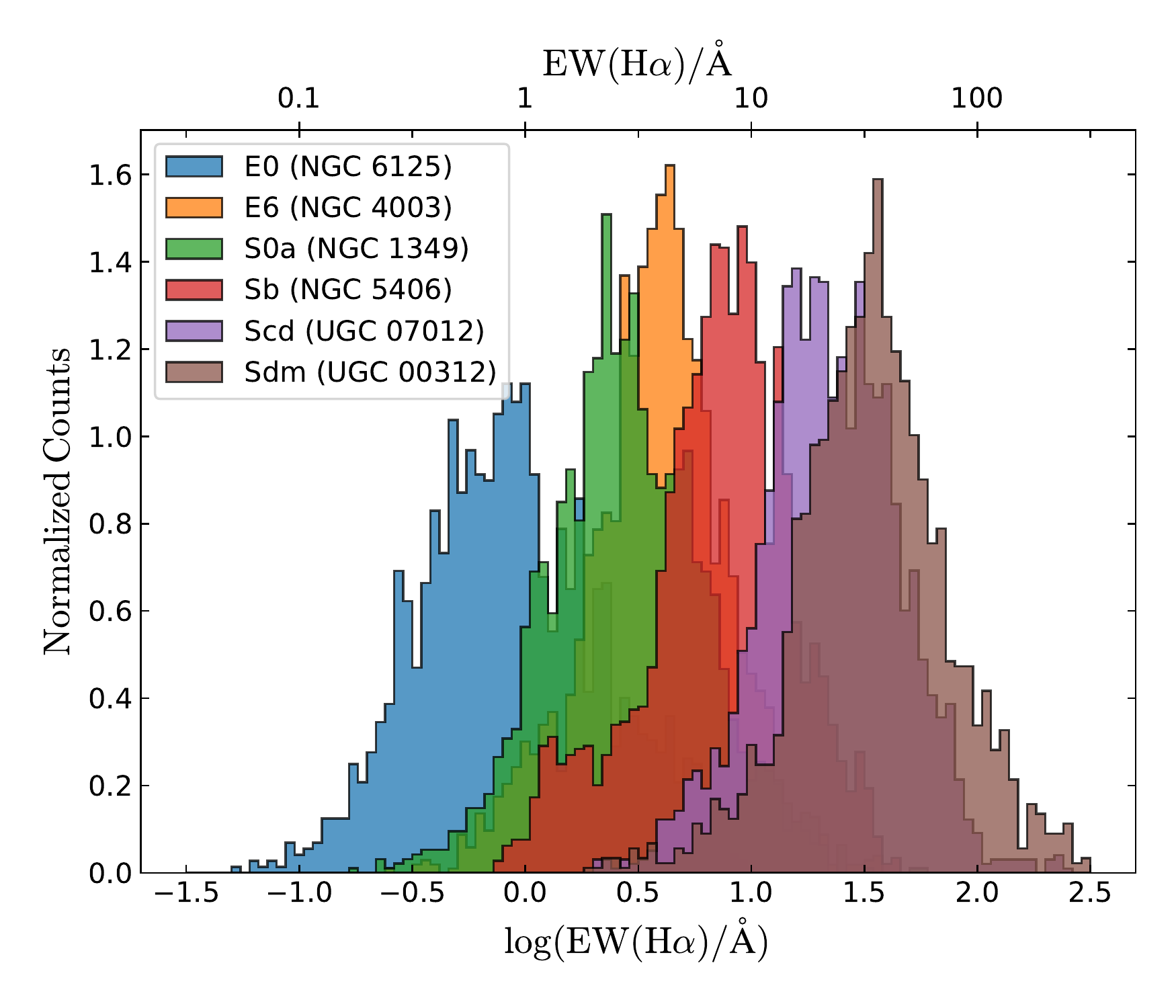}
    \caption{Normalized histograms of the \EWHa\ of DIG spaxels along the Hubble sequence. One representative galaxy was chosen for six morphological types: E0 (NGC\,6125), E6 (NGC\,1349), S0a (NGC\,4003), Scd (UGC\,07012), Sb (NGC\,5406) and Sdm (UGC\,00312).}
    \label{fig:EW_hist_galaxies}
\end{figure}
%%%%%%%%%%%%%%%%%%%%%%%%%%%%%%%%%%%%%%%%%%%%%%%%%%%%%%%%%%%%%%%%%%%%%%%%%%%%%%%%%%%%%%%%%%%%%%%%%%%%

%%%%%%%%%%%%%%%%%%%%%%%%%%%%%%%%%%%%%%%%%%%%%%%%%%%%%%%%%%%%%%%%%%%%%%%%%%%%%%
\begin{figure}
	\includegraphics[width=\columnwidth]{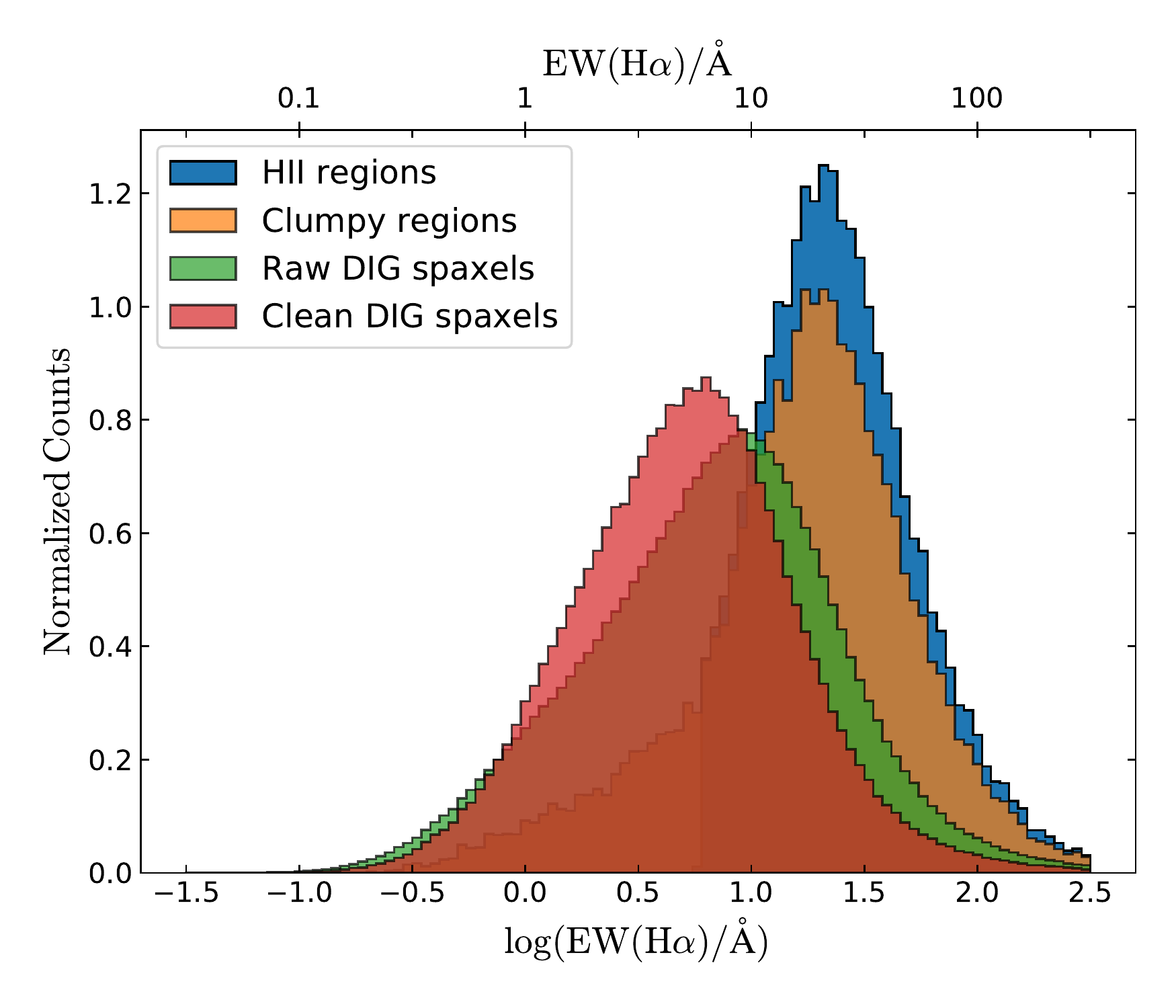}
    \caption{Normalized histograms of the \EWHa\ for the four different sub-groups of ionized regions described in the text: (i) the raw DIG spaxels (DIG spaxels for all galaxies); (ii) clean DIG spaxels (DIG spaxels for the clean galaxy sample); (iii) clumpy ionized regions (all those regions identified by \textsc{pyHIIexplorer}) and finally (iv) \HII\ regions (clumpy ionized regions selected as \HII\ regions, see \cref{sec:sel}).
    }
    \label{fig:hist_EW_all}
\end{figure}
%%%%%%%%%%%%%%%%%%%%%%%%%%%%%%%%%%%%%%%%%%%%%%%%%%%%%%%%%%%%%%%%%%%%%%%%%%%%%

%%%%%%%%%%%%%%%%%%%%%%%%%%%%%%%%%%%%%%%%%%%%%%%%%%%%%%%%%%%%%%%%%%%%%%%%%%%%%%%%%%%%%%%%%%%%%%%%%%%%%%%%%%%%%%%%%%%%%%%%%%%
\begin{figure*}
	\includegraphics[width=\textwidth]{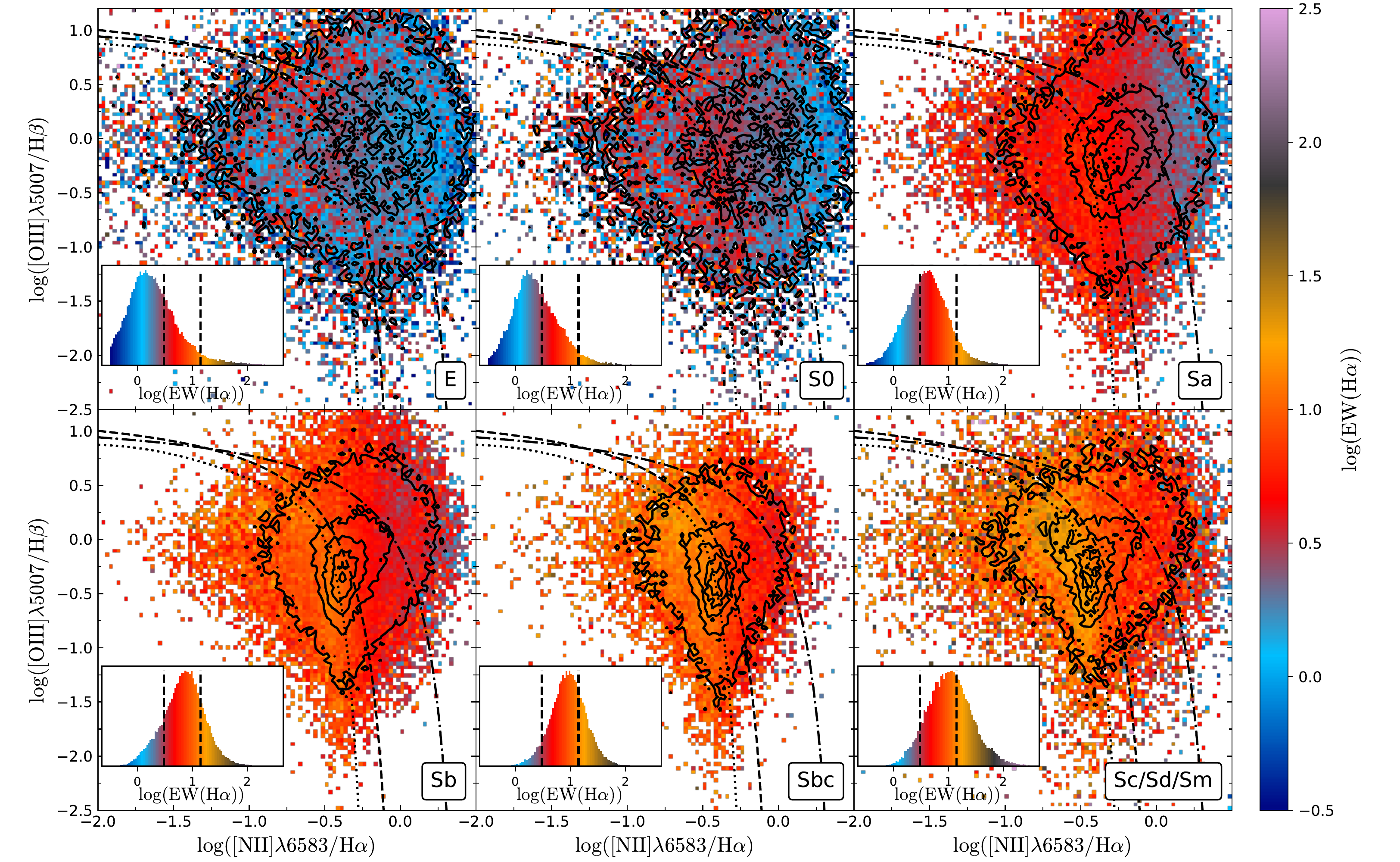}
    \caption{[OIII]$\lambda 5007$/H$\beta$ vs [NII]$\lambda 6583$/H$\alpha$ diagnostic diagrams for all the spaxels selected as  {\it clean DIG}, segregated by the morphology of their respective host galaxies. The color indicates the EW(H$\alpha$). The inset within each panel shows the EW(H$\alpha$) distribution of the corresponding spaxels. In these inset plots, we show the \EWHa$=3$\AA\ and \EWHa$=14$\AA\ as black dashed vertical lines. 
    The dashed, dot dashed and dotted lines represent the \citet{kauffmann03}, \citet{kewley01} and \citet{2008Stasinska_MNRAS391} demarcation curves, respectively. The density contours are similar to those presented in \cref{fig:EW_frac_star}
    }
    \label{fig:BPT_O3N2_diff_pts_morphs}
\end{figure*}
%%%%%%%%%%%%%%%%%%%%%%%%%%%%%%%%%%%%%%%%%%%%%%%%%%%%%%%%%%%%%%%%%%%%%%%%%%%%%%%%%%%%%%%%%%%%%%%%%%%%%%%%%%%%%%%%%

\cref{fig:EW_hist_galaxies} illustrates how the distribution of \EWHa\ for the DIG regions (spaxels excluded from the clumpy regions by \textsc{pyHIIexplorer}) correlates with respect to morphology (equivalent with the stellar ages).
Early-type galaxies (NGC\,6125, NGC\,1349 and NGC\,4003) tend to have low values of EW in comparison to late-type ones (NGC\,5406, UGC\,07012 and UGC\,00312) \citep{rosa14}. This is consistent wit results obteined by \citet{2018Lacerda_MNRAS474}. The late-type galaxies shows an equivalent-widths that correspond to ionization due to old stars or leaking/PSF-wings contamination associated with star-forming areas. Therefore, for early-type galaxies, the associated ionizing sources are predominantly old stars as expected since the star-formation is absent in these galaxies. Conversely, for late-type galaxies, both ionizing sources, old and young stars, should be observed. However, the number of spaxels with DIG associated with ionization by young stars should dominate over the number of them being ionized by old stars.
\footnote{A further study about the DIG sample and their distribution across the BPT diagram is present in \cref{app:DIG}}

In order to account for the effects of PSF-wings, we repeat the selection of DIG areas by enlarging the masked areas around \HII\ regions by one spaxel (i.e., one arcsec, $\sim$700pc at the redshift of the sample). However, although this procedure improves somehow the segregation, it does not guarantee a clean separation between the two ionizing sources. We repeat the procedure masking larger areas around the peak intensity of each \HII\ region. The larger the radius, the smaller is the area selected as DIG in each galaxy. Furthermore, the number of galaxies with any area classified as diffuse decreases. Thus, the statistics become poorer. Following this procedure, we are not able to obtain a clear segregation between the different ionizing sources even for very large masking radius (3-4 times the PSF FWHM), when most of the area within the FoV of the datacubes are masked.

As a final approach, we explore the properties of the DIG gas of the sample, considering all galaxies together, trying to derive an average correction that would be applied to each individual spaxel in each datacube. For doing so, we select all the DIG spaxels again for all the galaxies as described in \cref{sec:diff}, i.e., by selecting those areas outside the segregation maps used to select the \HII-region candidates. The green histogram in \cref{fig:hist_EW_all} shows the distribution of the \EWHa\ of the diffuse selected in this way (raw DIG). The distribution shows a peak centered on $\sim10$\AA, with an asymmetric shape, more tailed towards the low EWs values. This distribution corresponds to that of the mixed DIG or intermediate ionization as described by \citet{2018Lacerda_MNRAS474}, showing a clear contamination by different sources of ionization. For comparison purposes, the orange histogram in the figure shows the distribution of \EWHa\ for the clumpy ionized regions (as selected by \textsc{pyHIIexplorer}). As expected, it is clearly peaked towards higher EWs ($\sim$30\AA), although it presents a smooth but clear trend towards low EWs, so this indicates the presence of contamination by ionizing sources not compatible with \HII\ regions.

In an attempt to select just the best suitable spectra to characterize the DIG, we start selecting only those galaxies with high S/N diffuse emission. First,  we exclude those galaxies in which (i) the mean value of half of the emission lines from all DIG spaxels have a SNR value lower than 1.5, and (ii) if any average value of the emission lines present unrealistic values identified by eye for the DIG emission line fluxes (as a result of a bad fitting or a stellar population subtraction biases, or for regions at the edges of the hexagonal FoV of the data, in most of the cases).
The remaining galaxies comprise 30\% of the galaxy sample ($\sim 300$ galaxies, $\sim 2.3$ million of spaxels). Hereafter we will refer to the spectra of this galaxies sub-set as the ``{\it clean DIG}'' sample.
The effects of this selection are clearly appreciated in red histogram \cref{fig:hist_EW_all}. The distribution of \EWHa\ for the ``{\it clean DIG}'' sample has a peak now centered on $\sim5$\AA, showing a more symmetric shape. The new values for the EWs are more in agreement with an ionization dominated by old-stars. In order to investigate further the true nature of the ionizing source, we explore their distribution in the classical diagnostic diagrams.

\cref{fig:BPT_O3N2_diff_pts_morphs} shows the [OIII]$\lambda 5007$/H$\beta$ vs. [NII]$\lambda 6583$/H$\alpha$ diagnostic diagram  for all DIG spaxels. The spaxels were segregated by the morphology of their host galaxy in each panel, and the color represents the  mean \EWHa\ at each location. The \EWHa\ distribution for every group is plotted in an inset within each panel. There is a clear correlation between the location on the BPT diagram and the morphological type of the host galaxy. For early-type galaxies, the line ratios are mostly located in the intermediate area of the diagram (between the Kewley and Kauffmann curves). This region is dominated by a mixture of ionization by HOLMES/post-AGB and young-stars. In particular, these line ratios cannot be explained by star-forming models \citep{kauffmann03, kewley01, kewley06, stas06}. The distribution for late-types galaxies are more located below the demarcation curves. This zone is dominated by regions that are classically ionized by young stars and leaking photons from \HII\ regions.

Furthermore, the values of the \EWHa\ depends on the morphological type as well. The DIG in early-type galaxies tends to have low values of \EWHa. Conversely, late-type galaxies show higher values. Therefore, the \EWHa\ correlates with the morphology and with the position in the diagnostic diagrams. These results are expected by the correlation between the EW(H$\alpha$) and the ionizing source, according to \cite{2018Lacerda_MNRAS474}. However, for low \EWHa\ values, there are some spaxels below the Kauffmann's demarcation line: i. e., they are in the area assumed to correspond to ionization by young stars or post AGB \citep{mori16}.
This effect is considerably more severe before cleaning the sample based on the criteria indicated before, as shown in \cref{app:DIG} (\cref{fig:BPT_O3N2_DIG_EW}), highlighting the need for the applied selection.
In addition, there is a clear variation of the EW(H$\alpha$) values and the location on the BPT diagram along the Hubble sequence, in agreement with \cite{2018Lacerda_MNRAS474} and the examples are shown in \cref{fig:EW_hist_galaxies}.

We followed the same reasoning applied to select the \HII\ regions from the clumpy ionized areas (\cref{sec:sel}) to select those spaxels within the {\it clean DIG} sub-set corresponding to ionization due to old stars adopting the inverse criteria. We recall
that this criteria is based on the EW(H$\alpha$) and $f_y$ (see \ref{sec:sel}). Previously, we selected the regions with EW(H$\alpha$)$>6$\AA\ and $f_y>4\%$ as areas dominated by ionization due to young stars. On the contrary, to select DIG areas, we use the opposite constraints (EW(H$\alpha$)$<6$\AA\ and $f_y<4\%$) as spaxels ionized by old-stars. \cref{fig:EW_frac_star}, right-panel, shows the distribution of EW(H$\alpha$) against $f_y$ for the {\it clean DIG} spaxels. A direct comparison between the distributions in this diagram for clumpy ionized regions (left-panel) and DIG ones (right-panel) shows that most of the clumpy regions are already associated with areas consistent with being ionized by young stars. In contrast, most of the DIG ones are concentrated in areas associated with ionization by old-stars. We should note here that a fraction below 4\% of young stars is still reliable based on the inversion method adopted to derive the stellar population \citep[FIT3D][]{Pipe3D_I}. Different tests with similar codes indicate that below a 3\% the fraction of young stars is very unreliable \citep[e.g.,][]{2016Gonzalez_AA590}, and below this limit, all fractions should be considered just upper-limits \citep[e.g.,][]{2019Bitsakis_MNRAS483}. On the other hand, based on our criteria, the upper-right zone of the diagram for the DIG area corresponds, most probably, to contamination by leaking photons from \HII\ regions and the contribution of the PSF wings (that have similar properties in this diagram). The two other remaining quadrants are much less populated, and they are most probably contaminated by low-S/N DIG areas, both in the continuum and the emission line properties, that shift values from the other two ones which have a more clear physical interpretation.

\subsubsection{Decontaminating the HII regions by the DIG}
\label{sec:deco}

In the previous sections, we explore the distribution of the DIG and its nature. We selected a sub-set of galaxies from which it is possible to explore the properties of the DIG and define a procedure to determine which spaxels are ionized by old-stars. In this section, we describe how we derive the average properties of the emission line ratios for this final sub-set of spaxels (so far, for our dataset, the best possible representation of the DIG). Then, we show how we correct the emission line intensities of our catalog of \HII\ regions from this DIG contamination.

For a $j$-th emission line flux of the $i$-th \HII\ region, the corrected flux should be:

\begin{equation}
    \label{corr_eq}
    F_{\mathrm{corr}}^{i,j} = F_{\mathrm{obs}}^{i,j} - F_{\mathrm{diff}}^{i,j}
\end{equation}
where, $F_{\mathrm{obs}}^{i,j}$ is the observed emission line flux and $F_{\mathrm{diff}}^{i,j}$ is the emission line flux of the DIG to be corrected from. Thus, we need to estimate this later flux intensity.

In order to do so, we estimate the average flux intensity of all DIG regions included in the {\it clean} sample. First, for each spaxel $k$-th within this sample of DIG areas we normalize the flux of each $j$-th emission line by the flux of $\mathrm{H}\alpha$, defining the line ratio:
\begin{equation}
    \label{r_eq}
    R_{\mathrm{diff}}^{k,j} = \frac{F_{\mathrm{diff}}^{k,j}}{F_{\mathrm{diff}}^{k,\mathrm{H}\alpha}}
\end{equation}
Then, we derive the error weighted mean value of all these line ratios through all the {\it clean }sample of DIG spaxels, by using the formulae:

\begin{equation}
    \label{c_eq}
    C_{j} = \frac{\sum\limits_{k=1}^{k=n} \mathrm{w}_{k,j} R_{\mathrm{diff}}^{k,j}}{\sum\limits_{k=1}^{k=n} \mathrm{w}_{k,j}}
\end{equation}

where $\mathrm{w}_{k,j}$ is the square of the inverse of the error of the $j$-th emission line for the $k$-th spaxel, and $n$ is the total number of {\it clean} DIG spaxels.

Once derived $C_{j}$, i.e., the error weighted mean value of the line ratio with respect to
$\mathrm{H}\alpha$ for the $j$-th emission line for the DIG, we can calculate the DIG emission line flux of that line that would corresponds to a particular $i$-th HII-region, $F_{\mathrm{diff}}^{i,j}$, using the formulae:

\begin{equation}
    \label{f_eq}
    F^{i,j}_{\mathrm{diff}} = F^{i, \mathrm{H}\alpha}_{\mathrm{diff}} \times C_j,
\end{equation}

where $F^{i,\mathrm{H}\alpha}_{\mathrm{diff}}$ is the H$\alpha$ flux of the diffuse gas.
In principle, we do not have a direct way to estimate this flux. However, we can use the definition of the equivalent width, and determine the relation:

\begin{equation}
    \label{ew_eq}
    F_{\mathrm{diff}}^{i, \mathrm{H}\alpha} = \mathrm{EW_{\mathrm{diff}}^{i,H\alpha}}\times\rho^{i, \mathrm{H}\alpha}
\end{equation}

where $\rho^{i, \mathrm{H}\alpha}$ is the continuum flux density at the central wavelength of $\mathrm{H}\alpha$. We have direct access to this flux density as a direct dataproduct of the analysis by \textsc{Pipe3D}, which provides a version of the original datacube once subtracted the estimated contribution of the emission lines. These datacubes are used to estimate the EWs of all the emission lines, as explained in \citet{Pipe3D_II}. On the other hand, for the DIG (ionized by HOLMES or postAGBs), we consider that \EWHa$\approx1$\AA, based on theoretical estimations \citep{binn94,sarzi10,papa13,gomes16}. Indeed, this value is in agreement with our results shown in the right panel of \cref{fig:EW_frac_star}. Although the observational mean value is $\sim1.4$\AA, we prefer to use a conservative value, \EWHa$\approx1$\AA, to perform the DIG correction. It is important to note that this value is not related to the mean value shown in \cref{fig:hist_EW_all} since this distribution is dominated by the PSF wings contamination and the photons leaking, as indicated before.

Using \cref{ew_eq} and \labelcref{f_eq}, and the parameters $C_{j}$ estimated for each $j$-th emission line we estimate
$F_{\mathrm{diff}}^{i,j}$. Finally we decontaminate the contribution of the DIG for each emission line in each \HII\ region within the catalog using \cref{corr_eq}. The resulting values are stored in a new catalog of \HII\ regions corrected by DIG.

\subsubsection{Caveats to the adopted procedure}
\label{sec:diff_caveats}

We should note here that the adopted approach to derive the DIG contribution and its use to decontaminate line fluxes of \HII\ regions is based on different assumptions described below that may not be entirely valid. We have assumed that the only DIG contribution to correct for is the one produced by old-stars (HOLMES/post-AGBs), considering that the diffuse ionization by photon-leaking would produce similar emission line ratios. That assumption is far from been fully proved \citep[e.g.,][]{rela12}. Furthermore, \cite{asari19} recently studied the effects of a decontamination by DIG ionized by photon-leaking in the derivation of the oxygen abundance. Using SF galaxies from the MaNGA dataset \citep{manga}, they found that there is a difference up to 0.1 dex in the oxygen abundance calculated with the N2 index at the high metallicity end and low EW. However, there is no significant change in the oxygen abundances derived with other strong line indices (e.g., O23, O3N2, etc.). This agrees with our perception that this particular DIG does not alter reported line ratios significantly (besides maybe N2, with a small effect in any case). 

On the other hand, other contributions, like shock ionization, that could be considerably important at different scales \citep{2005Veilleux_ARAA43, carlos19}, are not considered here. Finally, we considered that all the DIG presents the same line ratios everywhere within a galaxy and for different galaxies. This is a very strong assumption that it is only valid at first order, as we have indeed shown in \cref{fig:BPT_O3N2_diff_pts_morphs}. The line ratios expected by HOLMES/post-AGB ionization cover a wide range of values \citep[e.g.,][]{gomes16,mori16}. However, our empirical exploration of these line ratios indicate a small degree of variation, almost compatible with the expected distribution due to their errors. In addition, we consider that
$\mathrm{EW_{\mathrm{diff}}^{H\alpha}}$ is 1\AA\ everywhere across the optical extension of galaxies and for all galaxies. This assumption, based on theoretical expectations, is empirically verified for the DIG of the earlier-type galaxies (\cref{fig:BPT_O3N2_diff_pts_morphs}). However, for later morphological types, the observed EWs are slightly larger. We assumed that the contamination by other sources of DIG and the wings of the PSF (from the adjacent \HII-region) alter the observed EWs of the DIG in these later-type galaxies. We acknowledge all these caveats to the adopted procedure. However, based on the performed experiments, we consider that this is the best approach we can adopt so far for the considered dataset. Being aware of possible improvements in the procedure, we provide with an uncorrected version of the \HII\ catalog together with the corrected one to allow the community to perform their evaluation on the issue of the DIG contamination.

\begin{figure*}
\includegraphics[width=\textwidth]{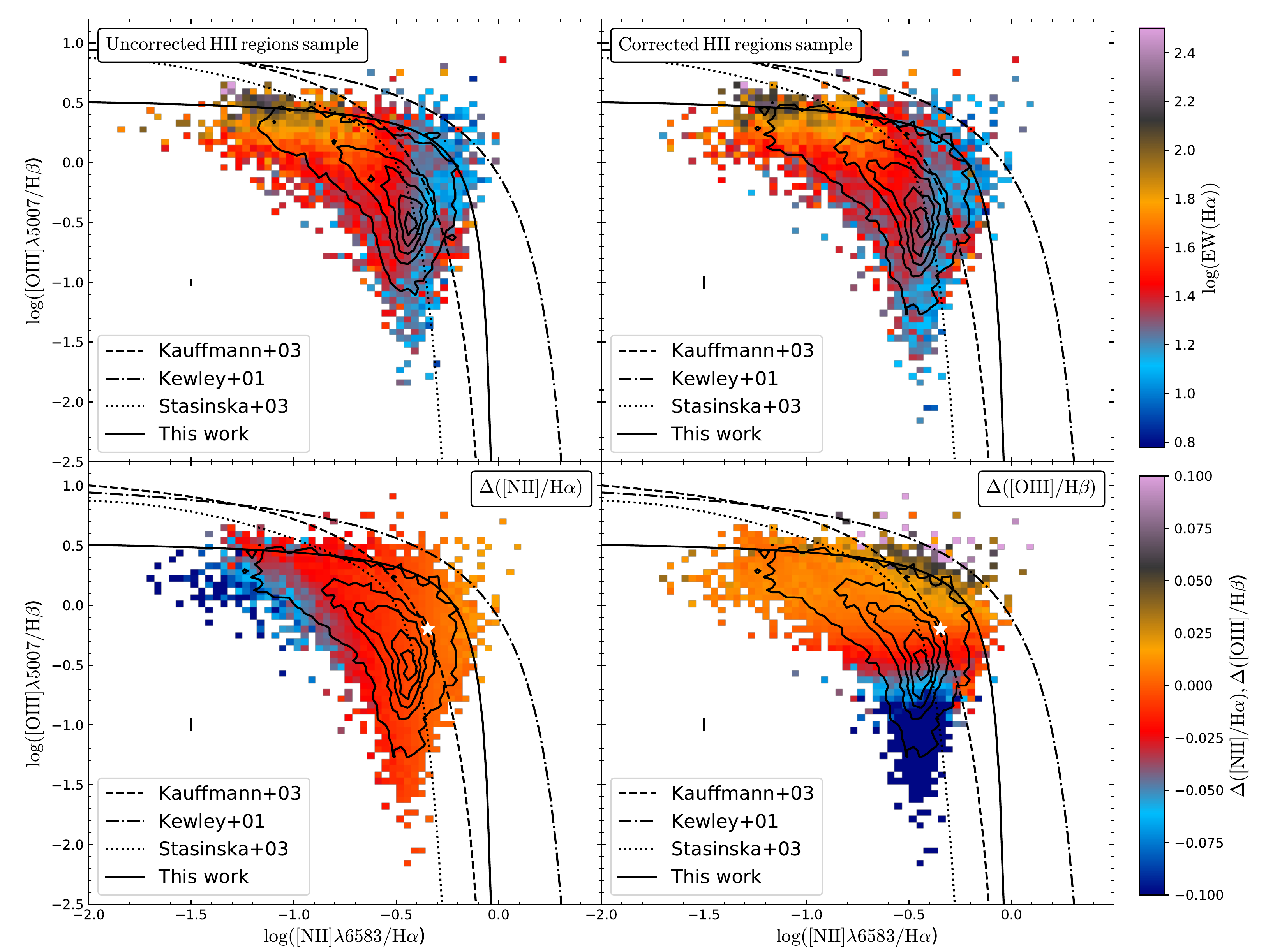}
    \caption{Distribution of the final sample of \HII\ regions along the BPT diagram. \textit{Left upper panel}: distribution without the correction of the contamination of the DIG. \textit{Right upper panel}: Distribution after correcting by the DIG contamination. In these panels, the color code represents the equivalent width of H$\alpha$.
    \textit{Left bottom panel}: Distribution of the corrected \HII\ regions with a color code representing the difference between the flux corrected and uncorrected for the $[\mathrm{NII}]/\mathrm{H}\alpha$ ratio ($\Delta \mathrm{NII}/\mathrm{H}\alpha$). \textit{Right bottom panel}: Distribution of the corrected \HII\ regions with a color code representing the difference between the flux corrected and uncorrected for the $[\mathrm{OIII}]/\mathrm{H}\beta$ ratio ($\Delta \mathrm{OIII}/\mathrm{H}\beta$). In these panels, the white star is the error weighted mean value for the corresponding line ratio.
    Notice that with the DIG correction the distribution has spread out slightly at the edges. This can be seen better on the density contours on upper panel and in the color gradient in the bottom panels. All lines represent the same demarcation curves described in \cref{fig:BPT_O3N2_diff_pts_morphs}. Also, the demarcation curve proposed in \cref{sec:demarc} is shown as a solid line.
    }
    \label{fig:BPT_O3N2_uncorr_corr}
\end{figure*}
%%%%%%%%%%%%%%%%%%%%%%%%%%%%%%%%%%%%%%%%%%%%%%%%%%%%%%%%%%%%%%%%%%%%%%%%%%%%%%%%%%%%%%%

\section{Results}
\label{sec:results}

\subsection{Effects of the DIG decontamination}
\label{sec:DIG_effect}

\cref{fig:BPT_O3N2_uncorr_corr} shows the distribution across the classical BPT diagnostic diagram for the final sample of \HII\ regions before and after applying the DIG correction (left-top and right-top panels). As can be seen, most of the uncorrected values are located below the maximum starburst envelope demarcation line \citep{kewley01}. This result is expected since the criteria to select \HII\ regions is based on the physical properties of this type of nebulae. After DIG correction, there is a change in the emission line fluxes and, consequently, in the respective equivalent widths, fluxes ratios and the distribution across the diagnostic diagram. The distribution stretches at the edges, covering a slightly wider range of line ratios. However, the average trend is very similar to the contaminated fluxes. Most of the decontaminated values remain in the SF zone in the diagnostic diagram. Besides, the EW(H$\alpha$) values decrease slightly on average, since the continuum flux density is constant along the DIG correction calculations (see \cref{sec:deco}), but the flux intensities have decreased slightly too. This effect is stronger for the \HII regions with lower \EWHa.

This result is somewhat expected. \cite{mast14} already shown the effects of the contamination by DIG on the flux ratios used in the explored diagnostic diagrams. Although our correction only takes into account a diffuse mainly powered by old stars and post-AGB (among other assumptions), our results show the same trend that they obtained with their methodology. Unfortunately, the spatial resolution of our data does not allow us to identify all diffuse gas components accurately and, hence, our data is not fully decontaminated from the complete DIG contribution. However, the proposed correction is enough to produce a small but noticeable change in the distribution of \HII\ regions along the BPT diagram.

\cref{fig:BPT_O3N2_uncorr_corr} shows the classical BPT diagram for the \HII\ regions sample corrected by DIG with (bottoms panels) a color code representing the difference between the flux corrected and the flux uncorrected for each line ratio in the diagram\footnote{Others diagnostic diagram are shown in the \cref{sec:app_demar_lines}}. As can be seen, the most significant  change in the emission line fluxes are in the edges of the distribution(upper left for $[\mathrm{NII}]/\mathrm{H}\alpha$ and lower right for $[\mathrm{OIII}]/\mathrm{H}\beta$). These results are in agreement with the change of the density points between the original distribution and the decontaminated one (upper panels of the same figure). Also, the error weighted mean value, $C_j$ (see \cref{sec:deco}), for the corresponding line ratios is shown as a white star. The value of $\Delta [\mathrm{NII}]/\mathrm{H}\alpha$ and $\Delta [\mathrm{OIII} ]/\mathrm{H}\beta$ depends on the position of $C_j$.

\subsection{An empirical demarcation line for HII regions}
\label{sec:demarc}

As indicated before, \cref{fig:BPT_O3N2_uncorr_corr} shows the classic diagnostic diagram for the entire sample of \HII\ regions (corrected and uncorrected by DIG). For comparison purposes, we include in \cref{fig:BPT_O3N2_uncorr_corr} the demarcation lines defined by \cite{kewley01} (dashed-dotted line), \cite{kauffmann03} (dashed line) and \cite{stas06} (dotted line). Since each demarcation line was derived on different assumptions, they do not provide the same classification (what is appreciated in the figure).

The most conservative demarcation line, in terms of minimizing the possible contamination by other ionizing sources different from \HII/SF-regions, is the one proposed by \cite{stas06} (dotted line in \cref{fig:BPT_O3N2_uncorr_corr}). This curve, like the one proposed by \cite{kewley01} (dot-dashed line in \cref{fig:BPT_O3N2_uncorr_corr}), defines the maximum envelop of line ratios derived by the location of pure SF regions based on photoionization models. The observed differences between both curves are due to the different models adopted in their derivations. Finally, an empirical demarcation line was derived based on a large sample of star-forming SDSS galaxies \citep[][in \cref{fig:BPT_O3N2_uncorr_corr})]{kauffmann03}. This line was defined by inspecting the BPT diagram visually as the line that better distinguishes the AGN from the star-forming galaxies. We should advise the reader that none of these demarcation lines should be interpreted as strict boundaries between different ionizing sources. As extensively discussed in previous articles \citep[e.g.,][]{sanchez18,sanchez19}, the demarcation lines should be interpreted as the upper-limit of line ratios that can be achieved due to ionization by massive/young OB stars
\citep{kewley01}. However, not all the ionizing regions located below that curves are associated with ionization by young stars per-se. As clearly shown in \cref{fig:BPT_O3N2_diff_pts_morphs}, the DIG produced by old-stars can be well bellowed any of those curves, in particular the one defined by \cite{kewley01}. The regions between the Kewley and Kauffmann curves have been extensively classified as ``mixed" or ``composite". It is true that for single aperture spectroscopic data that encompass a large portion of the optical extension of galaxies or for low-resolution IFS data (and/or particular inclinations), these line ratios could be reproduced by the combination of SF+AGN or SF+post-AGB ionizations \citep[e.g.,][]{mast14,davies16}. However, uncontaminated ionization by (i) certain types of \HII\ regions, like the N-enhanced/shock mixed ones described by \citet{Ho14} \& \citet{sanchez12a}; (ii) pure DIG produced by stars of certain ages \citep[e.g.,][]{mori16}; (iii) pure low-velocity shocks \citep[e.g.,][]{carlos19}, and even (iv) low-metallicity AGNs \citep[e.g.,][]{2017Stasinska_confE37} may be located in that area without requiring a mixed ionization.

In this study, we propose a new purely empirical demarcation line, that, like the other quoted ones, should be considered as an upper boundary for the involved line ratios due to ionization by young/massive stars. Based on the distribution of our sample of \HII\ regions, we select the envelop that includes 95\% of them as the boundary of ionized regions mainly powered by young stars. Then, by fitting these set of line ratios that defines the limit using the same functional form adopted in the literature \citep[e.g.,][]{kewley01, kauffmann03}, we obtain the following demarcation lines for the most commonly used diagnostic diagrams:

%%%%%%%%%%%%%%%%%%%%%%%%%%%%%%%%%%%%%%%%%%%%%%%%%%%%%%%%%%%%%%%%%%%%%%%%%%%%%%%%%%%%%%%%%%%%%%%%
\begin{figure*}
	\includegraphics[width=\textwidth]{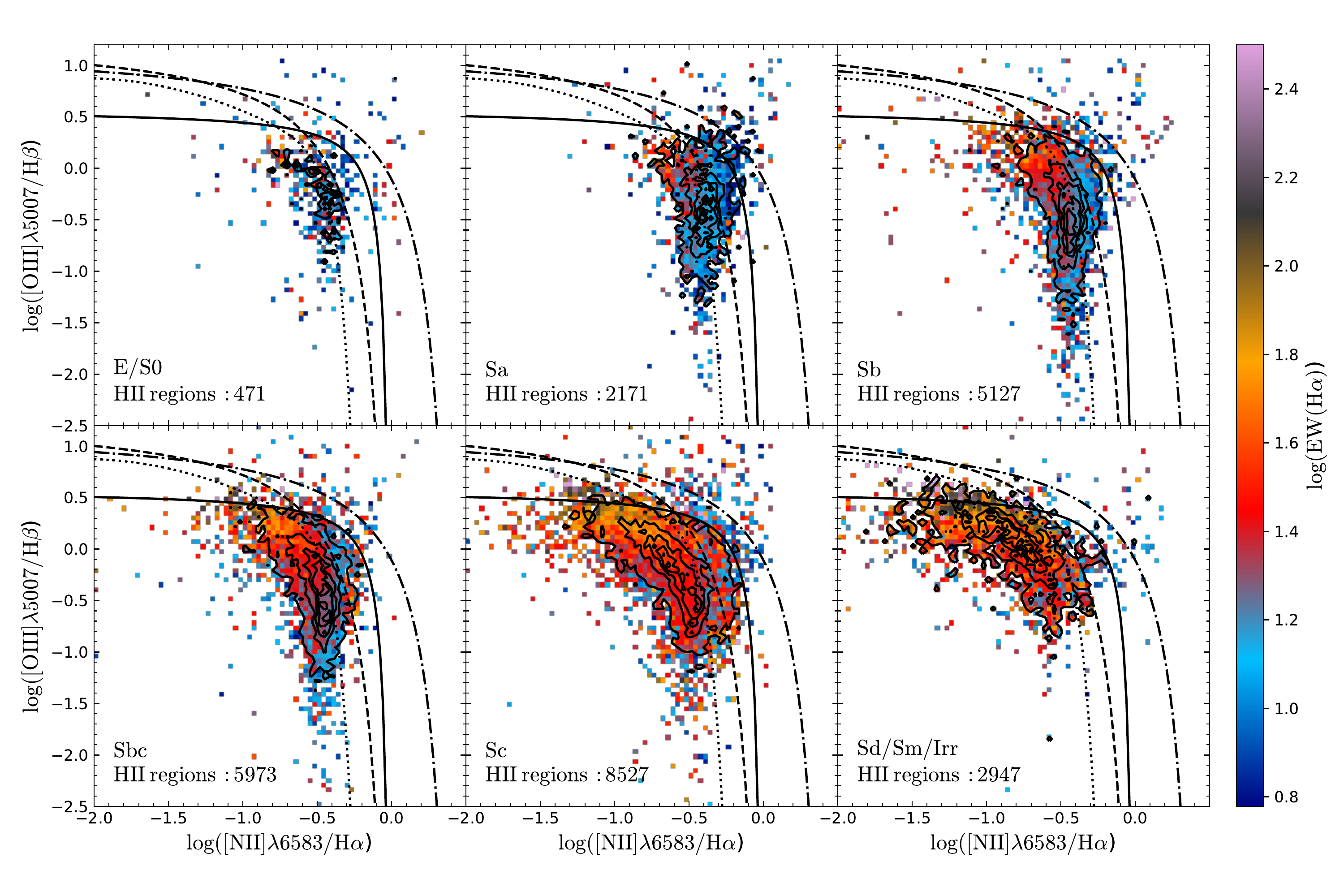}
    \caption{[OIII]$\lambda5007/$H$\beta$ vs [NII]$\lambda 6583/$H$\alpha$ diagram for the final sample of decontaminated \HII\ regions. The sample was divided in six groups according to morphological type of their galaxy host, and represented in a different panel comprising E/S0, Sa, Sb, Sbc, Sc and Sd/Sm/Irr galaxies the each one (from top-left to bottom-right). In each panel, the color code represent the mean \EWHa\ of \HII-regions. The density contours are similar to those presented in \cref{fig:EW_frac_star}. Also, the demarcation curve proposed in \cref{sec:demarc} is shown as a solid line.
    }
    \label{fig:BPT_O3N2_morph_seg}
\end{figure*}
%%%%%%%%%%%%%%%%%%%%%%%%%%%%%%%%%%%%%%%%%%%%%%%%%%%%%%%%%%%%%%%%%%%%%%%%%%%%%%%%%%%%%%%%%%%%%%%%

\begin{equation}
    \label{eq:O3N2_curve}
        \log([\mathrm{OIII}]/\mathrm{H}\beta) < 0.13/(\log([\mathrm{NII}]/\mathrm{H}\alpha) - 0.003) + 0.57
\end{equation}
\begin{equation}
    \label{eq:O3S2_curve}
        \log([\mathrm{OIII}]/\mathrm{H}\beta) < 0.04/(\log([\mathrm{SII}]/\mathrm{H}\alpha) + 0.012) + 0.58
\end{equation}
\begin{equation}
    \label{eq:O3O1_curve}
        \log([\mathrm{OIII}]/\mathrm{H}\beta) < 0.056 /(\log([\mathrm{OI}]/\mathrm{H}\alpha) + 0.40) + 0.61
\end{equation}

The curve for [OIII]/H$\beta$ vs [NII]/H$\alpha$ (\cref{eq:O3N2_curve}) is represented by a solid line in \cref{fig:BPT_O3N2_uncorr_corr}. The locations of the proposed demarcation lines for the remaining classical diagnostic diagrams are shown in the corresponding figures in \cref{sec:app_demar_lines} (\cref{fig:BPT_O3S2_new_curve} and \labelcref{fig:BPT_O3O1_new_curve}).

It is important to note that these new demarcation lines are based on a sample of \HII\ regions that were selected based on only two assumption: (i) a clumpy structure with a high contrast H$\alpha$ emission and (ii) an underlying stellar population comprising young stars (i.e., compatible with the assumed source of ionization).

\subsection{HII regions along the Hubble Sequence}
\label{sec:morph}

Early explorations on the properties of \HII\ regions have determined that later-type spirals present a larger number of \HII\ regions, distributed along (more) open spiral arms than earlier-type ones \citep{kennicutt88,kennicutt89a}. Furthermore, \citet{kennicutt89} and \citet{ho97} showed that \HII\ regions in the central areas of early spirals distinguish themselves spectroscopically from those in the disk by their stronger low-ionization forbidden emission. These results were confirmed by more recent studies using large IFS datasets \citep{sanchez12a,sanchez14b}. In particular, \citet{sanchez14b} demonstrated that the ionization properties of \HII\ regions strongly depends on the properties of the underlying stellar population, and, therefore, they present clear trends with the morphology as well. The enlarged \HII\ catalog presented here allows us to perform a more detailed exploration of this dependency.

\cref{fig:BPT_O3N2_morph_seg} shows the distribution of \HII\ regions across the BPT diagram segregated by the morphology of their respective host galaxies. The top left-hand panel shows the 254 regions found in E and S0 galaxies ($\sim 2\%$ of the total sample),
corresponding to 38 galaxies. Although early-type galaxies are frequently associated with the absence of star-forming regions, recent results indicate that a small fraction of them present \HII\ regions \citep[e.g.,][]{gomes16b}. They could be either remnants of a former disk or the result of a rejuvenation due to the capture of gas-rich galaxies or in falling gas \citep{gomes16b}. \cref{tab:hubble} shows the number of \HII\ regions along the different morphological types. Most of the \HII\ regions are located in late-type galaxies, with only $\sim$7\% of the \HII\ regions hosted by E/S0 galaxies. 
However, most of the \HII\ regions are not found in later spirals, but rather in Sb/Sbc galaxies, comprising 21\%/30\% of the total sample. On average, the number of \HII\ region per galaxy increases from earlier to later types up to Sc, spanning from a $\sim$7 regions in E/S0 galaxies to $\sim$50 in Sbc/Sc galaxies. Beyond that, for later spirals, the number decreases slightly. However, we cannot be completely sure if this decline is real or due to resolution effects. So far, the seminal studies by \citet{kennicutt88, 1998Kennicutt_ARAA36} indicate that the number of \HII\ regions should rise as later is the host, without any decline. The same articles indicate that \HII\ regions in late-type galaxies are significantly brighter than those of more early-types. 
However, the \HII\ regions are limited by the properties of CALIFA's data. Inevitably, we are losing most of the low luminosity \HII\ regions. As we will discuss later (see \cref{sec:disc}), in a future work, we will explore the methodology used in this work on data with a higher spatial resolution.

\begin{table}
\centering

\begin{tabular}{crrrrrr}

\hline
\hline
Morph. & N$_{gal_{T}}$ & N$_{gal}$ & \%$_{gal}$ & N$_{\HII}$ & \%$_{\HII}$ & 
$\left <\mathrm{N}_{\HII}\right >$\\ 
\hline
E & 163 & 44 & 4.8 & 190 & 0.7 & 4.3 \\
S0 & 105 & 46 & 5.0 & 415 & 1.5 & 9.0 \\
Sa & 135 & 121 & 13.1 & 2572 & 9.5 & 21.3 \\
Sb & 137 & 132 & 14.3 & 5598 & 20.6 & 42.4 \\
Sbc & 113 & 112 & 12.1 & 6274 & 23.1 & 56.0 \\
Sc & 172 & 172 & 18.6 & 8917 & 32.9 & 51.8 \\
Sd & 65 & 65 & 7.0 & 2259 & 8.3 & 34.8 \\
Sm & 21 & 21 & 2.3 & 633 & 2.3 & 30.1 \\
Irr & 10 & 10 & 1.1 & 214 & 0.8 & 21.4 \\

\hline
\hline
\end{tabular}
\caption{HII regions by morphology.
$1)$ Morphology of the host galaxies. $2)$ Number of galaxies. $3)$ Number of galaxies with detected \HII\ regions. $4)$ Fraction of galaxies with \HII\ regions detected. $5)$ Number of \HII\ regions per morphological type of host galaxy. $6)$ Fraction of \HII\ regions. $7)$ Average number of \HII\ regions per galaxy 
}
\label{tab:hubble}
\end{table}

The relation between the distribution of \HII\ regions across the BPT diagram and the morphology of their host galaxy uncovered by \citet{sanchez14b} is appreciated in \cref{fig:BPT_O3N2_morph_seg}. The few \HII\ regions found in early-type galaxies are located around the central region of the BPT diagram without any significant trend, maybe due to the low number  of statistics. The trend becomes more evident when exploring the distributions from early-spirals (Sa) towards later ones, with \HII\ regions shifting along the classical sequence of these objects \citep[e.g.,][]{osterbrock89}.

The location in the BPT diagram of a \HII\ region is primarily related to the ionization conditions of the nebulae \citep[e.g.,][]{kewley02,mori16}. Thus, a priori, it is not expected a connection between this location and the morphological properties of the galaxy that host the \HII\ regions, which is what we see precisely in \cref{fig:BPT_O3N2_morph_seg}. The low number of \HII\ regions in early-type galaxies, which star-forming activity is finished \citep[or almost finished,][]{2008Stasinska_MNRAS391, 2009Blanton_ARA&A47}, was expected. However, the nature of the observed trend within the BPT diagram with the Hubble sequence, besides the clear increase in the number of \HII\ regions, is less obvious. We can discard strong contamination by diffuse ionized gas as the primary driver for the observed distribution. First, the selection of clumpy ionized regions, together with the implemented cuts (\cref{sec:sel}), guarantees that we have indeed selected regions in which OB stars dominate ionization. Second, as shown in \cref{sec:DIG_effect}, this contamination shrink somehow the range of covered parameters, but it does not blur the observed trends completely. Finally, we have performed a detailed DIG subtraction, so its contribution is minimized.

%%%%%%%%%%%%%%%%%%%%%%%%%%%%%%%%%%%%%%%%%%%%%%%%%%%%%%%%%%%%%%%%%%%%%%%%%
\begin{figure*}
	\includegraphics[width=\textwidth]{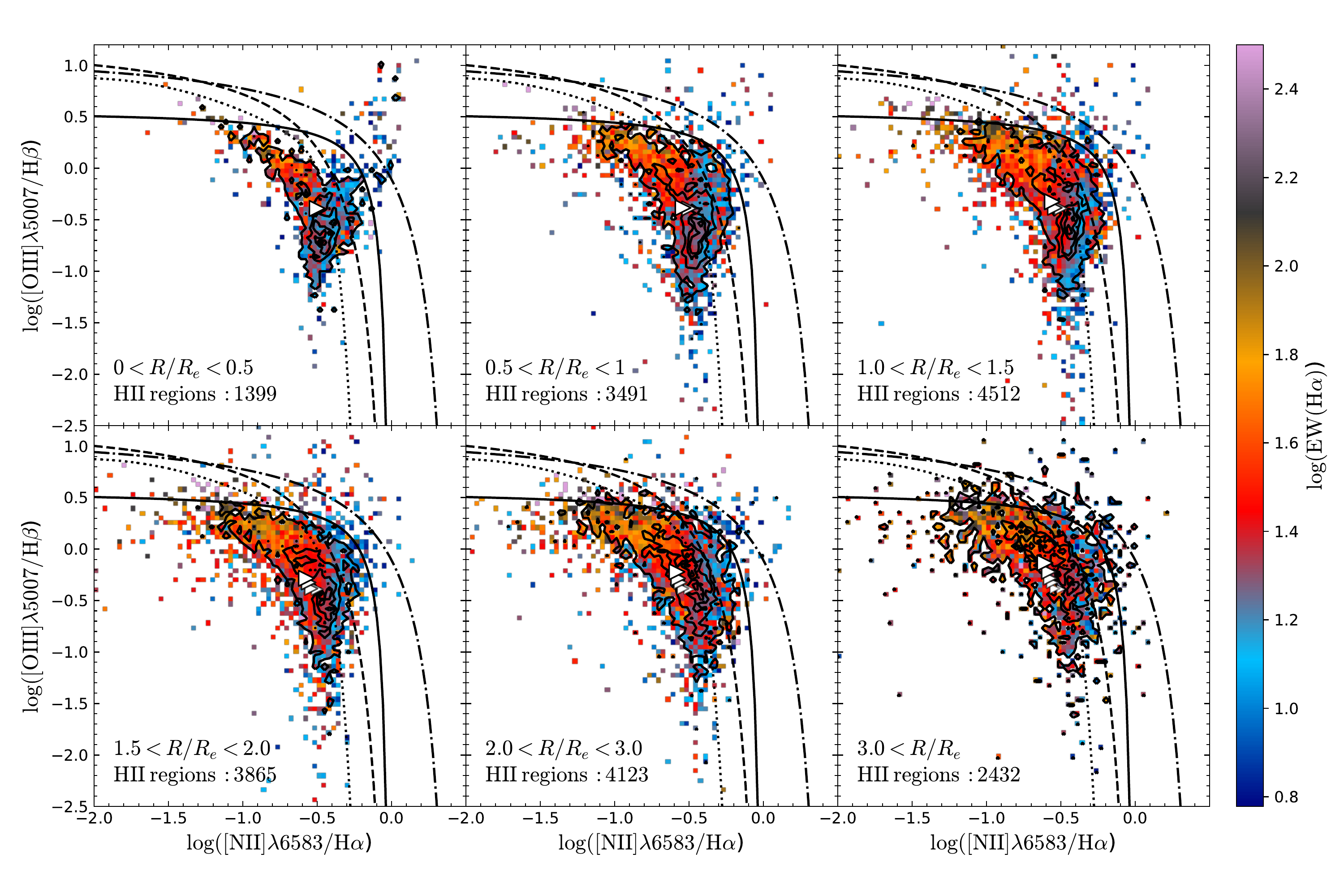}
    \caption{[OIII]$\lambda5007/$H$\beta$ vs [NII]$\lambda 6583/$H$\alpha$ diagram for the \HII\ regions. The sample was divided into six groups according to their galactocentric distance, normalized to the effective radius: i) $R/R_e>0.5$ (top left hand); ii) $0.5<R/R_e<1$ (top center panel); iii) $1<R/R_e<1.5$ (top right hand ); iv) $1.5<R/R_e<2$ (bottom left hand); v) $2<R/R_e<3$ (bottom center panel), vi)  $R/R_e>3$ (bottom right hand). The density contours are similar to those presented in \cref{fig:EW_frac_star}. The color code indicates the mean \EWHa\  value in the respective bin. The white triangles in each panels represent the mean value of the [OIII]/H$\beta$ and [NII]/H$\alpha$ ratios. Also, the previous values of the each panels are show in the next panels as transparent white triangles. All lines represent the same demarcation curves described in \cref{fig:BPT_O3N2_diff_pts_morphs}. Also, the demarcation curve proposed in \cref{sec:demarc} is show as a solid line.
    }
    \label{fig:BPTs_O3N2_dist_seg}
\end{figure*}
%%%%%%%%%%%%%%%%%%%%%%%%%%%%%%%%%%%%%%%%%%%%%%%%%%%%%%%%%%%%%%%%%%%%%%%%%

Assuming that OB stars indeed ionize all the selected regions, the observed trend should be the consequence of variations between these stars (and the surrounding nebulae). For the most late-type galaxies (Sd, Sm and irregular), the location (upper left end on the diagram) is normally interpreted as the presence for low metallicity stars with high ionization strengths. Meanwhile, for earlier-spirals (Sa or Sb), the location (lower right end on the diagram) is linked to higher metallicity stars with lower ionization strengths (these results are also found and discussed by S\'anchez in press). Finally, in intermediate galaxy types, the \HII\ regions extends within the two regimes. Despite of the significant effect of the aging of the ionizing stars (that would impose certain variations within the observed trends), the only obvious conclusion is that we are observing a trend driven by a change in the metallicity, that induces a change in the ionization strength due to an anticorrelation\footnote{Recent studies were found that in some galaxies exists a positive correlation between both parameters \cite[e. g.,][]{2018Poetrodjojo_MNRAS479, 2018Thomas_ApJ856}} between both parameters \citep[e.g.,][]{dopita16cal,mori16, 2016Pilyugin_MNRAS457}. If so, the observed trend implies a gradation with morphology in the stellar metallicity (in particular, the oxygen abundance).

\subsection{HII regions along galactocentric distances}
\label{sec:dist}

In the previous section, we explore the distribution of the \HII\ regions segregated by the morphological type of their host galaxies across the diagnostic diagrams. In this section, we explore the distribution of the \HII\ regions across the BPT for different galactocentric distances. The distance was directly derived from the \HII\ regions catalog, which comprises the physical properties of the region by their location with respect to the center of the galaxies, in addition to many other parameters. Based on these coordinates, we derive the galactocentric distance, correcting them by the inclination of each galaxy and position angle, and normalizing by the effective radius.

%%%%%%%%%%%%%%%%%%%%%%%%%%%%%%%%%%%%%%%%%%%%%%%%%%%%%%%%%%%%%%%%%%%%%%%%%%%%%%%%%%%%%%%%%%%%%%%%%%%%%%%%%%%%%%%%%%%
\begin{figure*}
	\includegraphics[width=\textwidth]{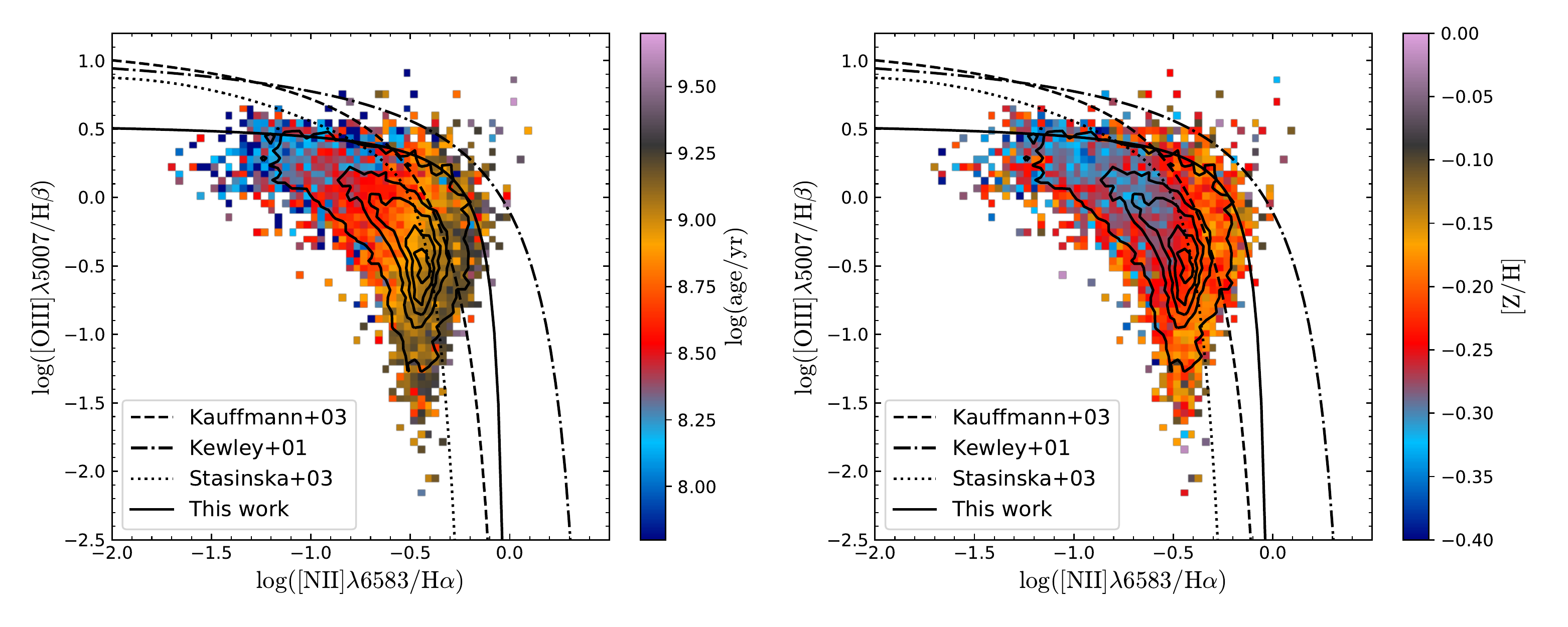}
    \caption{[OIII]$\lambda5007/$H$\beta$ vs [NII]$\lambda 6583/$H$\alpha$ diagram for the \HII\ regions in our final catalog. \textit{Left panel}: the color indicates the average value of the luminosity-weighted age of the underlying stellar population. \textit{Right panel}: the color indicates the average value of the luminosity weighted metallicity of the underlying stellar population. The density contours are similar to those presented in \cref{fig:EW_frac_star}. In both panels, the dotted, dot-dashed, dashed and solid lines are the demarcation curves used in \cref{fig:BPT_O3N2_uncorr_corr}.
    }
    \label{fig:stellar_properties}
\end{figure*}
%%%%%%%%%%%%%%%%%%%%%%%%%%%%%%%%%%%%%%%%%%%%%%%%%%%%%%%%%%%%%%%%%%%%%%%%%%%%%%%%%%%%%%%%%%%%%%%%%%%%%%%%%%%%%%%%%%%%

\cref{fig:BPTs_O3N2_dist_seg} shows the distribution of \HII\ regions across the BPT diagram segregated into different galactocentric distances, ranging from the smallest to the largest distances. The distribution changes clearly with the distance to the center of the galaxy.
Those regions towards the center ($R/R_e<0.5$) are mostly located at the right-end of the BPT diagram, at a very similar location were early spirals \HII\ regions are found (\cref{fig:BPT_O3N2_morph_seg}). On the other hand, the regions at larger distances ($R/R_e>0.5$) follow the classical trend for these objects (e.g., \citealt{osterbrock89}), with a shift towards the upper-left range of the distribution as the galactocentric distances increases. This result indicates that the ionization conditions of the \HII\ regions depends on its location in the galaxy. This result is observed with the mean lines ratios (white triangles) at different galactocentric distances. Earlier results already found that the classical \HII\ regions are those within the disk of spiral galaxies \citep[e.g.,][]{diaz89}. However, towards the center of galaxies, they present stronger low-ionization forbidden emission lines \citep[e.g.,][]{kennicutt89, ho97}, shifting them towards the right-hand range of the diagram close to the demarcation curve. \citep[e.g.,][]{sanchez12a}. These studies suggested that this may be due to the contamination by an additional source ionization, as diffuse emission or shocks. However, the data in \cref{fig:BPTs_O3N2_dist_seg} is already corrected by the contribution of DIG and the distribution of \HII\ regions remains at the same location in the BPT diagram. In general, we confirm the results by \citet{sanchez14b} in this regard, where the observed trends were already reported using a more limited number of \HII\ regions (over a more limited number of galaxies) and without performing a DIG correction.

A plausible explanation for the enhanced low ionization forbidden line ratios in central \HII\ regions could be the contamination by shocks, as suggested by \citet{ho97}. The presence of shocks contamination in these regions was first suggested by \citet{stauffer81} and explored to explain the differences in the disk and central \HII\ regions by other authors \citep[e.g.,][]{peimbert92}. Other authors do not report a clear difference between regions found at different galactocentric distances \citep[e.g.,][]{veilleux87}. Indeed, although there is a clear trend towards the right-hand regime of the BPT diagram for \HII\ regions in the center of galaxies, not all central \HII\ regions are located at that extreme end. Therefore, if shocks are the source of the extra ionization, it is not present in all \HII\ regions. A plausible explanation is that remnants of recent super-novae explosions associated with the star-formation events that ionized the observed region could contaminate the line ratios, deviating them from the expected location due to pure photoionization by OB stars. Under this scenario, our results indicate that central \HII\ regions are more prompt to this contamination. We will explore this possibility in forthcoming studies. As a first approach, models with shock \citep[like MAPPINGS models,][]{2017Sutherland_ApJS229} could be used to study the contribution of shocks, using, for example, the grid recently published by \citet{Alarie19}.

\subsection{HII regions and the underlying stellar population}
\label{sec:age_met}

In the previous, sections we confirmed the results by \citet{sanchez14b}, indicating that the location of an \HII\ region within the BPT diagram is associated with the galactocentric distance and the morphology of the host galaxy. These authors connected the reported trends with the properties of the underlying stellar population (i.e., not the ionizing population) at the location (and host) of the \HII\ regions. It is known that the bulk of the stellar population of the center of disk galaxies present similarities with that of early-type galaxies \citep[e.g.,][]{rosa14,godd15,sanchez18}. Moreover, they present radial gradients in both the stellar ages and metallicities \citep[e.g.,][]{rosa14}, the sSFR \citep[e.g.,][]{2016Gonzalez_AA590,sanchez18} and the gas content \citep[e.g.,][]{utomo17,sanchez18}, and the overall star-formation histories \citep{rgb17}. Thus, the ionization conditions seem to be connected somehow with the local stellar evolution within the considered galaxy.

\begin{figure*}
	\includegraphics[width=\textwidth]{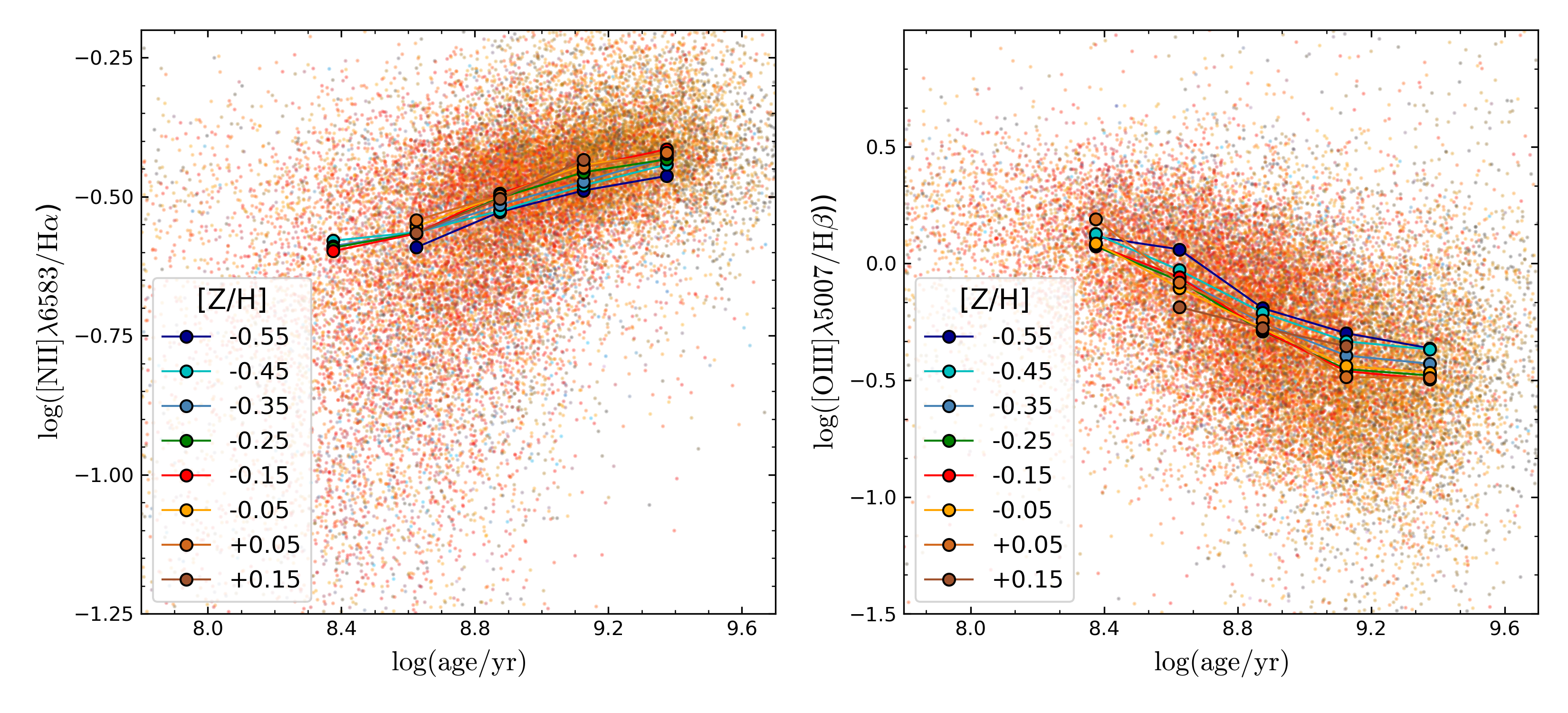}
    \caption{Distribution of the [NII]$\lambda 6583/$H$\alpha$ line ratio (\textit{left panel}) and [OIII]$\lambda5007/$H$\beta$ (\textit{right panel}) along the luminosity-weighted age of the underlying stellar population for the \HII\ regions in our catalog, color-coded by the [Z/H] of the underlying stellar population. The solid circles represent the average line-ratios within bins of 0.25 dex in log(age/yr) for a fixed range of metallicites (binned in ranges of 0.1 dex). The average [Z/H] corresponding to each color is included in the inset of each panel. All circles within the same metallicity range are linked by a solid lines and plotted with the same color (that codifies [Z/H]).
    }
    \label{fig:N2_O3_vs_Age_ZH}
\end{figure*}

\cref{fig:stellar_properties} shows the distribution of \HII\ regions across the BPT diagram with a color code representing properties of underlying stellar populations. The left panel shows the luminosity-weighted ages and the right panel shows the metallicity of the underlying stellar population. The properties of the underlying stellar populations at the location of each \HII\ regions was extracted from the dataproducts provided by {\sc Pipe3D}, as described in \cref{sec:Pipe3D}, following \cite{sanchez13} and \cite{sanchez14}. As can be seen in \cref{fig:stellar_properties}, the location of an \HII\ region in the BPT diagram depends on both stellar parameters. The regions with a younger underlying stellar population are located in the upper-left area of the diagram. Otherwise, the regions with an older underlying stellar population are located in the right-end of the diagram.

Similarly, the regions with poor metal underlying stellar populations are found in the upper left area, while the regions with metal-rich underlying stellar populations are mostly located at the right end of the diagram. It is important to remark that we are not talking about properties of the ionizing population. The reported ages and metallicities correspond to the total stellar population at the location of each \HII\ region. This result is consistent with the one obtained by \citet{sanchez14b}, where similar trends were reported.

The trends along the BPT diagram with morphology and galactocentric distance reported in previous sections are naturally explained by the distributions shown in \cref{fig:stellar_properties}. Early-type galaxies (and the center of all galaxies) are dominated by old and metal-rich stellar populations \citep{rosa14, godd15, 2015Sanchez_AA573A}. In general, and therefore, the \HII\ regions on those galaxies are located at the right end of the distribution for these objects. On the contrary, late-type galaxies (and the outer areas of all galaxies) are mostly dominated by younger and more metal-poor stellar populations. Therefore, the \HII\ regions on those galaxies are located in the left-end of the distribution.

\subsection{Imprints of the galactic evolution in the ionization}
\label{corr}

As described in the previous section, the location in the diagnostic diagrams depends on the underlying stellar population properties (age and metallicity). Thus, the BPT diagram exhibits a connection between the stellar population properties and the line ratios. We explore this connection in here in a more quantitative way.

\cref{fig:N2_O3_vs_Age_ZH} shows the distribution of [\ion{N}{ii}]/H$\alpha$ line ratio (left-hand panel) and the [\ion{O}{iii}]/H$\beta$ line ratio (right-hand panel) along the luminosity-weighted age of the corresponding underlying stellar population at the location of each \HII\ region. The color code represents the stellar metallicity. As can be seen in the figure, there exists a trend between the lines ratios and the stellar properties. Indeed, the distributions present a correlation coefficient of $r=0.50$ (0.63) for low metallicities, [Z/H]$\sim$-0.55 dex, and $r=0.46$ (0.48) for high metallicities, [Z/H]$\sim$0.15 dex, for the [\ion{N}{ii}]/H$\alpha$ ([\ion{O}{iii}]/H$\beta$) line ratio.
In general, the older and more metal-rich underlying stellar population is related to a higher (lower) values of the [\ion{N}{ii}]/H$\alpha$ ([\ion{O}{iii}]/H$\beta$) line ratio.

In order to illustrate more clearly the correlation described above, we averaged the values in bins of 0.1 dex in [Z/H] and 0.25 dex in log(age). The results are shown in \cref{fig:N2_O3_vs_Age_ZH}, where every colored circle is the mean value of the corresponding line ratio and age in each bin for a fixed metallicity (coded by the color). 
It is clear the relation of the line ratios on both parameters when explored the binned data.

To characterize the reported trends, we fit both line ratios with a linear
combination of both properties of the stellar population properties (age and [Z/H]). We consider that with the current dispersion adopting a more complex functional form would not provide a better characterization. The best linear regression fitting for the [\ion{N}{ii}]/H$\alpha$ line ratio was:

\begin{equation}
    \begin{split}
        \log\left(\frac{[\ion{N}{ii}]}{\mathrm{H}\alpha}\right) = & 0.20_{\pm0.02} \log(\mathrm{age/yr})\\ & + 0.06_{\pm0.02}\mathrm{[Z/H]} -2.32_{\pm0.26}
    \end{split}
\end{equation}

The standard deviation of the residual of the line ratio, once subtracted the best-fitted model, is $\sim 0.166$ dex. Compared with the standard deviation of the original distribution of values ($\sim 0.194$ dex), the standard deviation was reduced\footnote{The difference was calculated by quadratic difference.} by a factor $\sim 52\%$. Therefore, the [\ion{N}{ii}]/H$\alpha$ largely depends on the underlying stellar population. For [\ion{O}{iii}]/H$\beta$, the best linear regression fitting with the properties of the underlying stellar population is:

\begin{equation}
    \begin{split}
        \log\left(\frac{[\ion{O}{iii}]}{\mathrm{H}\beta}\right) = & -0.58_{\pm0.01}\log(\mathrm{age/yr}) \\ & - 0.17_{\pm0.02}\mathrm{[Z/H]} + 4.94_{\pm0.12}
    \end{split}
\end{equation}

The residual between the line ratio observed and the best fit presents a dispersion of $\sim 0.348$ dex. Again, compared with the initial dispersion $\sim 0.382$ dex, the standard deviation was decreased by a factor $\sim 41\%$. Like in the case of the previous line ratio, this modeling clearly illustrates the reported relation. However, despite the reduction of the dispersion, the line ratios still present a wide range of values. Other properties influence the observed line ratios besides the underlying stellar properties. In particular, the physical conditions of the nebulae (i. e. electronic density, ionization parameter, metallicity, the geometry of the ionized gas, etc.) must indeed influence the line ratios.

\section{Discussion and Conclusions}
\label{sec:disc}

In this study, we presented a new catalog of \HII\ regions extracted from the eCALIFA ($+$ pCALIFA) $+$ PISCO sample. This new catalog comprises information of $\sim 28,000$ \HII\ regions, including the flux of the most important emission lines in the optical range (between 3745-7200\AA) and the corresponding properties of the underlying stellar population. It is important to remark that we based the selection of \HII\ regions on basically two assumptions: (i) they should appear as a clumpy structure in the H$\alpha$ emission line map (see \cref{sec:det}) and (ii) the underlying stellar populations should be compatible with the definition of \HII\ region (see \cref{sec:sel}).

However, as can be seen in \cref{fig:seg_map_contours}, the identification algorithm is not perfect. We are missing small clumpy regions and, also, we can not resolve those regions too close one each other. This is a direct consequence of the spatial resolution of the CALIFA data. As noticed by \cite{mast14}, with a low spatial resolution, the smallest ionized regions can not be segregated and they are erroneously assigned to bigger adjacent ones or confused with the diffuse. This is why almost all identified ionized regions have a similar size in our catalog. Thus, an important fraction of small ionized regions is lost in the final sample. This has a direct effect on the calculation of some properties, like the luminosity function \citep[e.g.,][]{sanchez12b}, that cannot be well recovered in the low luminosity regime. 

Furthermore, the wings of the PSF limit our ability to segregate among the different ionizing sources. First, they affect our ability to segregate between adjacent \HII\ regions, as the boundaries of each clumpy ionized region are not well defined at the current coarse resolution. Therefore, the segregation algorithm developed to identify the ionized regions could erroneously associate pixels that contain certain emission coming from the PSF wing of any adjacent (brighter) region to a fainter one, modifying the estimated values of the derived parameters. Second, due to these wings, the DIG is polluted by the emission of any adjacent \HII\ region too. Finally, \HII\ regions could be contaminated by the emission of other possible ionizing sources (not only DIG but shocks or AGNs) as well. This may somehow affect the physical properties calculated from the emission line fluxes \citep[][]{mast14,zhang16}.

To minimize the effects of this contamination, we present a new approach to correct the emission line fluxes of the \HII\ regions by the contribution of the DIG. As we mention in \cref{sec:diff}, the DIG emission was calculated by considering all the spaxels in between those selected as being part of a \HII\ region by our code. Again, spatial resolution plays an essential role in the proposed DIG correction. If we could identify exactly the limits of each ionized region (or model its shape), we would be able to identify clearly the DIG and, hence, make a proper correction irrespectively of the source of ionization of this diffuse (hot, old stars and leaking photons). Unfortunately, as indicated before and discussed in \cref{sec:diff_caveats}, we have an important contamination by ionization from the \HII\ regions themselves due to the coarse resolution and relatively large PSF wings. This contribution does not allow us to distinguish the DIG powered by leaking photons, for instance: the physical properties of this contamination, like EW(H$\alpha$) and line ratios, are too similar to that of adjacent \HII\ regions. Hence, we have corrected only the DIG powered by old, hot stars, which indeed is the one that may produce the strongest effects in the line ratios. This result is supported by the study of \cite{asari19}. They proposed a DIG correction where its contribution is calculated using only SF galaxies from the MaNGA dataset, i.e., a DIG of higher \EWHa. There is a small but not significant change between the metallicities calculated by the corrected emission lines and the uncorrected ones for most of the explored calibrators (beside N2, which presents a 0.1 dex offset). Besides, the physical spatial resolution of the data analyzed in \cite{asari19} is  nearly twice worst than the one included here (the FWHM$\sim 0.8$ kpc for CALIFA and $\sim 2.5$ kpc for MaNGA) what makes any correction by the DIG more complicated and with a stronger expected effect.
To compare the contribution of all DIG types, we required observations with a high spatial resolution. However, the DIG correction present here, although not perfect, is an overall improvement with respect to not applying any correction at all.

It is clear that better spatial resolutions will facilitate the segregation of the clumpy ionized regions. It will improve the detection of smaller (thus, fainter) \HII\ regions, the separation between adjacent regions, the definition of the DIG (and its different sources), and the overall decontamination of the DIG. New generation datasets, like the ones provided by MUSE \citep[e.g.,][]{laura18}, and SITELLE \citep[e.g.,][]{2018Rousseau-Nepton_MNRAS477}, would require a re-evaluation of the proposed procedure, both for the detection of the \HII\ regions and for the decontamination by the DIG. 

A good example of the capabilities of new instruments is the study by \citet{laura18}. They used MUSE observations of 102 galaxies, selected from the \textit{All-weather MUse Supernova Integral-field Nearby Galaxies} survey \citep[AMUSING; ][L\'opez-Cob\'a, submitted]{2016Galbany_MNRAS455}, to analyze the shape of oxygen abundance profiles. For doing so, they obtained a catalog of \HII\ regions using {\sc HIIexplorer}. The total number of regions detected was 14,345. The spatial resolution of their data, FWHM$\sim 0.8''$ (corresponding to the natural seeing) compared to the current value of $2.5''$ for CALIFA allows them to identify a number equivalent to $\sim$65\% of our current catalog in a number of galaxies that comprises just $\sim$10\% of the ones analyzed here.

On the other hand, SITELLE provides observations where the natural seeing also limits resolution. Using this instrument, \citet{2018Rousseau-Nepton_MNRAS477} identified 4285 \HII\ region candidates in a single galaxy (NGC 628). However, the methodology used for the identification of \HII\ regions generates several spurious ionized regions. In comparison, using PINGS data (of similar projected resolution than CALIFA), and {\sc HIIexplorer}, \citet{sanchez12b} recovered 373 \HII\ regions in the same galaxy. Using the same data, we recovered 243 \HII\ regions (from 244 ionized regions detected). The difference between the number of recovered \HII\ regions by \citet{sanchez12b} and us is due to the input parameters used in {\sc HIIexplorer} and {\sc pyHIIexplorer}. In \cite{sanchez12b}, the input parameters are listed.

In comparison, their minimum flux is lower than the value used in our calculations. However, they report that only 282 are \HII\ regions with a good quality spectra and good estimation of the H$\beta$ flux. Meanwhile, the 243 \HII\ regions that we report in here are the final number after applying the quality check test described in the text. Taking into account these details, we consider that the number of \HII\ regions found in both studies is consistent and the differences are clearly justified. Thus, the use of SITELLE represents a theoretical  improvement of a factor 10 in the detection of these ionized sources. 

Despite the limitations due to the coarse spatial resolution, there are still significant advantages in the current dataset. MUSE does not cover the blue wavelength range of the optical spectrum below $\lambda<$4650\AA.  Therefore fundamental emission lines are missing in the data provided by that instrument for galaxies at low redshift. Moreover, the information provided for the stellar populations is limited since the critical wavelength range around the 4000\AA\ break is not covered. With SITELLE the information for the stellar population is even less reliable due to the instrumental technique that the features of the continuum strongly. Finally, CALIFA allows us to present the current exploration for a statistically significant sample of galaxies, representative of the population at the nearby universe. Thus, the current \HII\ regions catalog is not (a priori) biased by morphology, mass or other galaxy properties. However, it is recognized that the possible properties derived from weak emission lines could drive to bias results, e.g., electron temperature derived from [\ion{O}{III}]$\lambda 4363$. This is a direct effect of the inherent weakness of these emission lines and/or possible contamination from other sources (e.g., sky emission lines). However, the properties derived for weak emission lines could be calculated well enough from the \HII\ regions.

We study the relationship between the location on the BPT diagram of the distribution of DIG-corrected \HII\ regions and (i) the morphology of the host galaxies; (ii) the galactocentric distances of the \HII\ regions; and (iii) the properties of the underlying stellar population. First, we demonstrate that the distribution of \HII\ regions based on our simple selection criteria are located in the star-forming area of the diagnostic diagrams (without this being a pre-requisite or part of the selection process). 
We found that the location in the BPT diagram depends strongly on the value of the EW(H$\alpha$).
This result was already reported in previous studies \citep[e.g.,][]{2015Sanchez_AA573A, 2018Lacerda_MNRAS474}.
However, in comparison with them, we demonstrate that the result still holds when using a larger statistical sample of \HII\ regions covering a more extensive range of physical properties and ionization conditions. 

We clearly showed that the location in the diagnostic diagram depends on the morphological type of the host galaxy and the galactocentric distance.
Early type galaxies, in general, massive and metal-rich, have \HII\ regions located in the lower-right end of the diagnostic diagram. Contrarily, in late-type galaxies, less massive and metal-poor, the distribution moves towards the upper-left zone of the diagrams. Thus, the change of location in the diagram is a consequence of the change of the stellar population properties along the Hubble sequence. \citet{2015Gonzalez_Degaldo_AA581} already noticed that the metallicity, age and other properties of stellar populations depends on the morphological type of galaxies. Indeed, they proved that the average stellar metallicity increase with the morphology.

Similarly, we found that the distribution in the BPT diagrams depends on the galactocentric distance. The most straightforward explanation is the same as the one described for the differences reported by morphological type. The center of galaxies is populated by older and more metal-rich stellar populations, while the outer regions are populated by younger and more metal-poor ones \citep[e.g.,][S\'anchez et al. in press]{2015Gonzalez_Degaldo_AA581}. This radial gradient is also evident in the oxygen abundance, which decreases with the galactocentric distances, too \citep[e.g.,][]{sanchez12b, sanchez14}.

This result is related to the inside-out scenario for the formation of galaxies. Other studies \citep[e.g.,][]{perez13, ibarra16, rgb17} demonstrated the radial dependence on the star formation history and the timescales of the gas infall. Therefore, the chemical evolution of the interstellar medium is affected by the star formation history at different distances given rise to the metallicity gradients on galaxies.

All the discussed results confirm that the locations within the BPT diagrams are tightly related to the properties of the underlying population, as already suggested by \citet{2015Sanchez_AA573A}. This is clearly shown in \cref{fig:stellar_properties}. Moreover, we demonstrate that the line ratios [NII]$\lambda 6583/$H$\alpha$ and [OIII]$\lambda5007/$H$\beta$ present a clear correlation with both the ages and metallicities of the underlying stellar populations (\cref{fig:N2_O3_vs_Age_ZH}). This relation is stronger for the [\ion{N}{ii}$\lambda 6583$]/H$\alpha$ line ratio. This result is somehow expected since this line ratio is widely used as a proxy of the oxygen abundance. The previous generation of stars enriched the chemical composition, oxygen and other element abundances in the ionized regions. Hence, the oxygen abundance of the ISM naturally correlates with the metallicity of the younger stars, provided that the radial mixing is not strong \citep{laura16,2015Sanchez_AA573A}.

On the other hand, there is a correlation with the age of the stellar population due to the chemical enrichment processes in galaxies. It is well known that, in disk-dominated galaxies, stars are older and more metal-rich in the central regions than the stars in the outer regions of the disk \citep{2015Gonzalez_Degaldo_AA581, 2014Sanchez-Blazquez_AA570, perez13}. Thus, metallicity and its relation with the differential star-formation history within galaxies and at different locations in them \citep[e.g.,][]{rgb17} is the most plausible explanation for the observed trends.

The other line ratio, [\ion{O}{iii}]$\lambda 5007$/H$\beta$, also shows a correlation with both analyzed stellar parameters (age and [Z/H]). However, since this ratio is less sensitive to oxygen abundance and more sensitive to variations in the geometry, density and temperature of the ionized nebulae, it presents a weaker correlation than that found for  [\ion{N}{ii}]$\lambda 6583$/H$\alpha$.

Although other authors already noticed some of the results discussed above, we present them, for the first time, for a statistically well-defined sample of \HII\ regions in the nearby Universe, several times larger than those analyzed previously. We demonstrate that those results do not depend strongly on the possible contamination by DIG since we have applied a correction to that contribution.
Moreover, we present a new demarcation line defined with this sample that is consistent with well known previous ones, although it is totally unbiased regarding the involved line ratios.

The current catalog of \HII\ regions and the full set of emission lines discussed in this article are publicly available. To our knowledge, this is the largest catalog of \HII\ region properties, derived from the broadest range of galaxies of different morphologies and masses. With the analysis presented in this study, we have demonstrated that this sample is suitable to perform further explorations on the physical properties of these nebulae (in particular for those that do not require detailed knowledge of the size and absolute luminosity). In forthcoming studies, we will present the main patterns of physical properties for these \HII\ regions.

\section{Acknowledgements}

We thank the referee for the helpful comments that have improved the quality of this article. We are grateful for the support of a CONACYT grants CB-285080, FC-2016-01-1916 and CB2015-254132, and funding from the DGAPA/PAPIIT-IA101217, DGAPA/PAPIIT-IN100519 and DGAPA/PAPIIT-107215 (UNAM) projects.

This study uses data provided by the Calar Alto Legacy Integral Field Area (CALIFA) survey (\url{http://califa.caha.es/}), it based on observations collected at the Centro Astron\'omico Hispano Alem\'an (CAHA) at Calar Alto, operated jointly by the Max-Planck-Institut f\"ur Astronomie and the Instituto de Astrof\'isica de Andaluc\'ia (CSIC).

L.G. was funded by the European Union's Horizon 2020 research and innovation programme under the Marie Sk\l{}odowska-Curie grant agreement No. 839090.

This research makes use of python (\url{http://www.python.org}), of Matplotlib \citep[a framework to the creation of scientific plots for Python,][]{2007Hunter_matplotlib}, Astropy \citep[a Python package for Astronomy,][]{astropy:2013, astropy:2018} and SciPy \citep[a Python-based ecosystem of open-source software for mathematics, science, and engineering,][]{Scipy}

%%%%%%%%%%%%%%%%%%%%%%%%%%%%%%%%%%%%%%%%%%%%%%%%%%


\bibliographystyle{mnras}
\bibliography{references,CALIFAI} % if your bibtex file is called example.bib

%%%%%%%%%%%%%%%%%%%%%%%%%%%%%%%%%%%%%%%%%%%%%%%%%%

%%%%%%%%%%%%%%%%% APPENDICES %%%%%%%%%%%%%%%%%%%%%

\appendix

\section{Demarcation lines and effects of DIG decontamination on others diagnostic diagrams}
\label{sec:app_demar_lines}

\begin{figure*}
	\includegraphics[width=\textwidth]{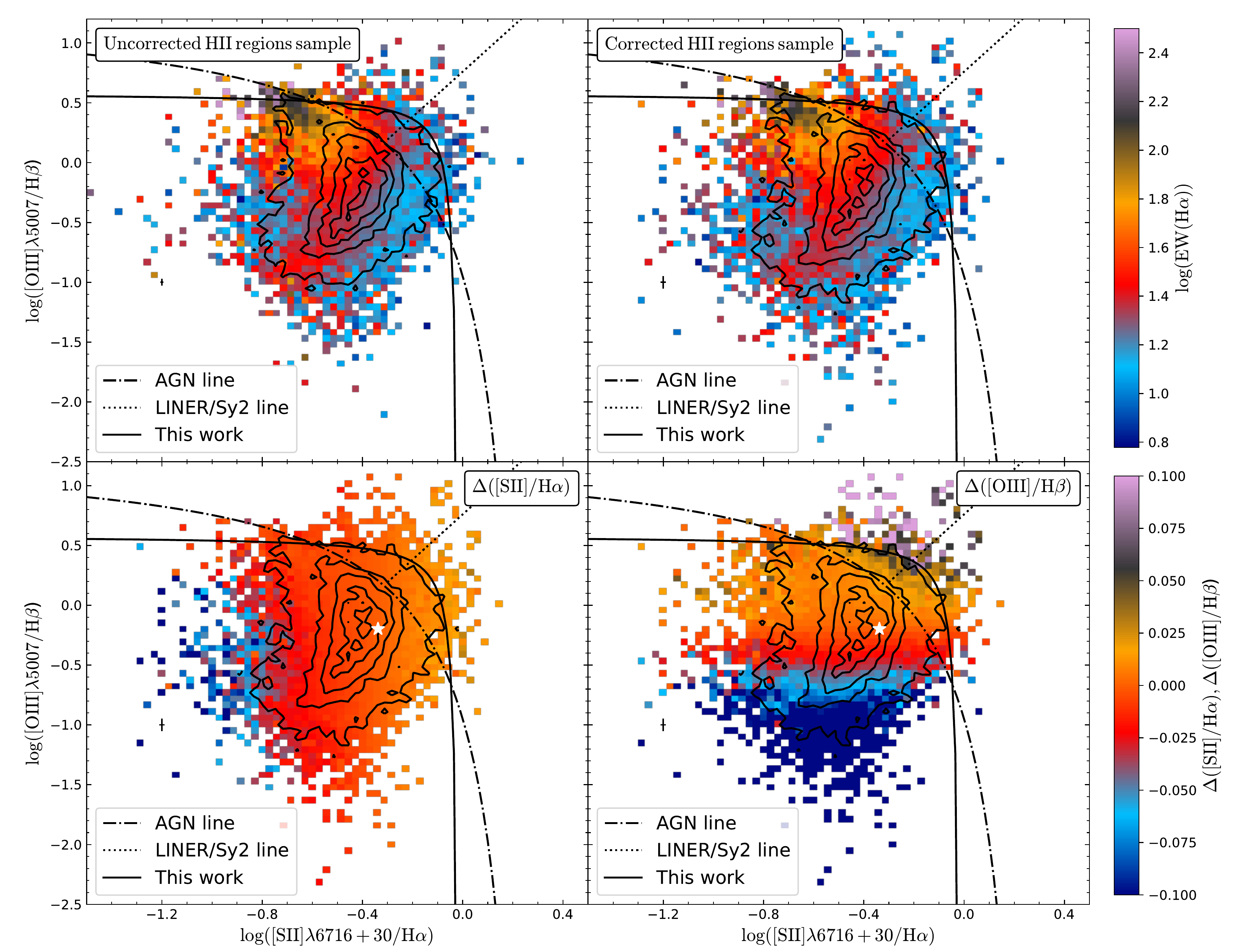}
    \caption{Distribution of \HII\ regions along the [OIII]$\lambda 5006$/H$\beta$ vs [SII]$\lambda 6716 + 30$/H$\alpha$ diagnostic diagram. \textit{Left panel}: distribution without DIG correction. \textit{Right panel}: distribution with DIG correction. The color code indicates equivalent width of H$\alpha$. The most frequently used demarcation curves for this diagram are shown \citet{kewley06}. The density contours are similar to those presented in \cref{fig:BPT_O3N2_uncorr_corr}
    }
    \label{fig:BPT_O3S2_new_curve}
\end{figure*}

\begin{figure*}
	\includegraphics[width=\textwidth]{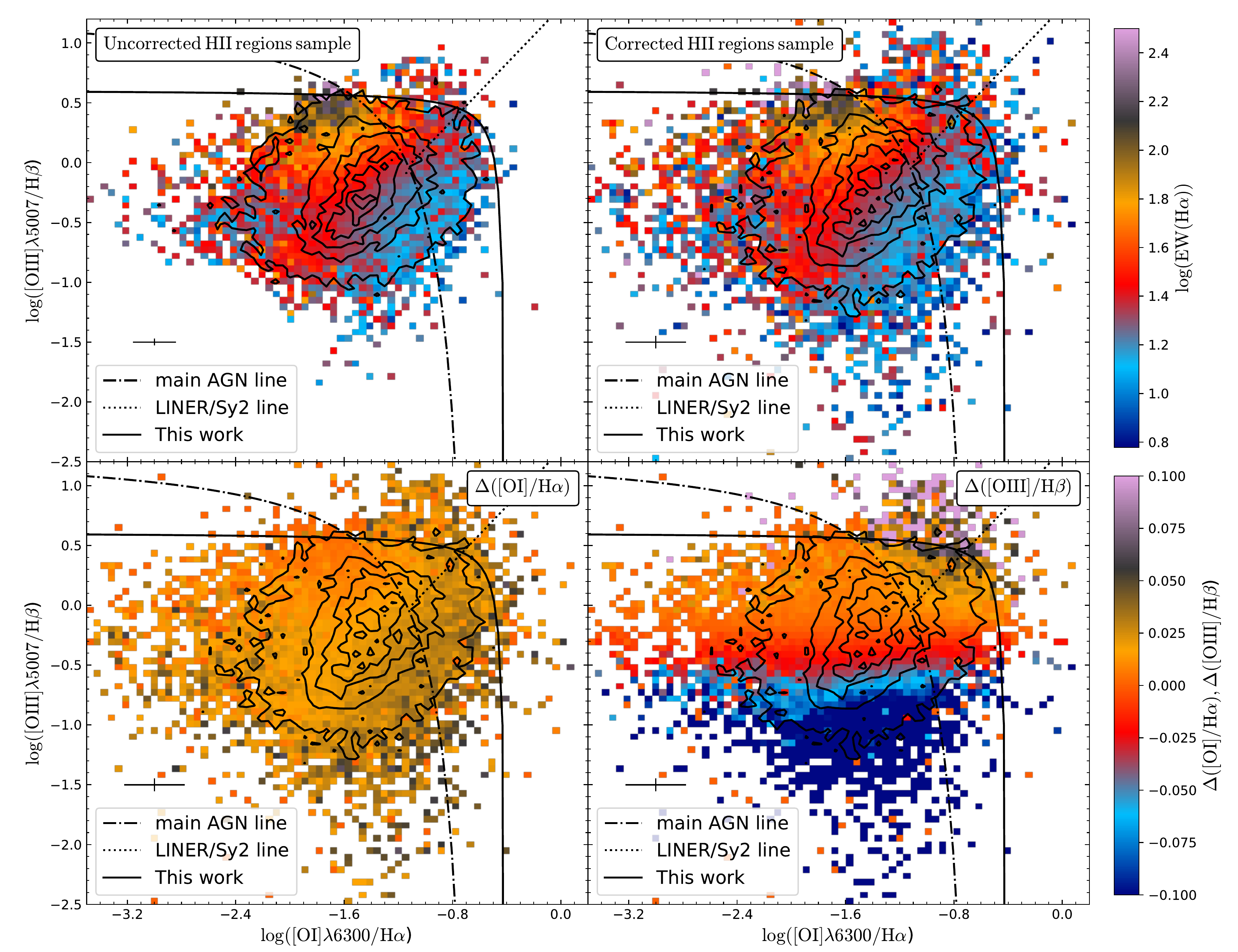}
    \caption{Similar figure as the one presented in \cref{fig:BPT_O3S2_new_curve} for the [OIII]$\lambda 5006$/H$\beta$ vs [OI]$\lambda 6300$/H$\alpha$ diagnostic diagram.}
    \label{fig:BPT_O3O1_new_curve}
\end{figure*}

In \cref{sec:demarc}, we present new demarcation lines for \HII\ regions for the most frequently used diagnostic diagrams: Equations (\ref{eq:O3N2_curve}), \labelcref{eq:O3S2_curve}, and \labelcref{eq:O3O1_curve}. The upper panels on \cref{fig:BPT_O3S2_new_curve} and \labelcref{fig:BPT_O3O1_new_curve} shows the [OIII]$\lambda 5006$/H$\beta$ vs [SII]$\lambda 6716 + 30$/H$\alpha$ and [OIII]$\lambda 5006$/H$\beta$ vs [OI]$\lambda 6300$/H$\alpha$ diagnostic diagrams for the DIG uncorrected and corrected \HII\ regions sample respectively (see \cref{fig:BPT_O3N2_uncorr_corr}). The most frequently used demarcation curves \citep{kewley06} for each diagram, together with our proposed ones are plotted too.

The distribution of regions in the different diagnostic diagrams corresponds to the expected one from previously published results. However, we should note that our selection did not use any of those line ratios, but a rather simple set of assumptions involving the shape of the region and the stellar population properties of the ionized regions (\cref{sec:sel}).

Despite the overall agreement with previous selections, there are some differences. In the [OIII] vs. [OI] diagram, we observe that the $\sim$14\% of our selected \HII\ regions lie in the AGN zone according to the classical demarcation lines. After the DIG correction, this fraction remains $\sim$13\%. Thus, the differences with classical selection criteria cannot be attributed to this effect.

In both diagrams, the new demarcation curves present a different shape than the classical ones, the SF zone defined by the new curves is higher than the classical ones. As we include regions with an EW(H$\alpha$) $\sim$ 6\AA, our curve could include some regions that have a mixture in the ionizing sources, according to \citet{2018Lacerda_MNRAS474}. Therefore, the classical lines provide a more conservative classification scheme to identify star-forming regions; meanwhile, the new curve defines a classification scheme that includes star-forming regions with the largest contribution of other ionizing sources than the classical ones.

In \cref{sec:DIG_effect}, we present the effects of DIG correction across the BPT diagram (\cref{fig:BPT_O3N2_uncorr_corr}). The bottom panels on \cref{fig:BPT_O3S2_new_curve} and \labelcref{fig:BPT_O3O1_new_curve} show the changes between the corrected and the uncorrected fluxes across their corresponding diagnostic diagrams ([OIII]$\lambda 5006$/H$\beta$ vs [SII]$\lambda 6716 + 30$/H$\alpha$ and [OIII]$\lambda 5006$/H$\beta$ vs [OI]$\lambda 6300$/H$\alpha$) respectively. As was mentioned before, there is a clear dependence between the location on diagnostic diagrams and the difference of fluxes, $\Delta([\mathrm{SII}]/\mathrm{H}\alpha)$, $\Delta([\mathrm{OIII}]/\mathrm{H}\beta)$, $\Delta([\mathrm{OI}]/\mathrm{H}\alpha)$. Unfortunately, for [OI]$\lambda 6300$/H$\alpha$ ratio, the coefficient's order, $C_j$, is less than the associate error order; hence, the dependence is not clear in the diagnostic diagram. This issue, as we discussed before, may be improved by using data with a better resolution.

\section{Quality of the HII regions sample}
\label{sec:ratios_check}
\begin{figure*}
	\includegraphics[width=\columnwidth]{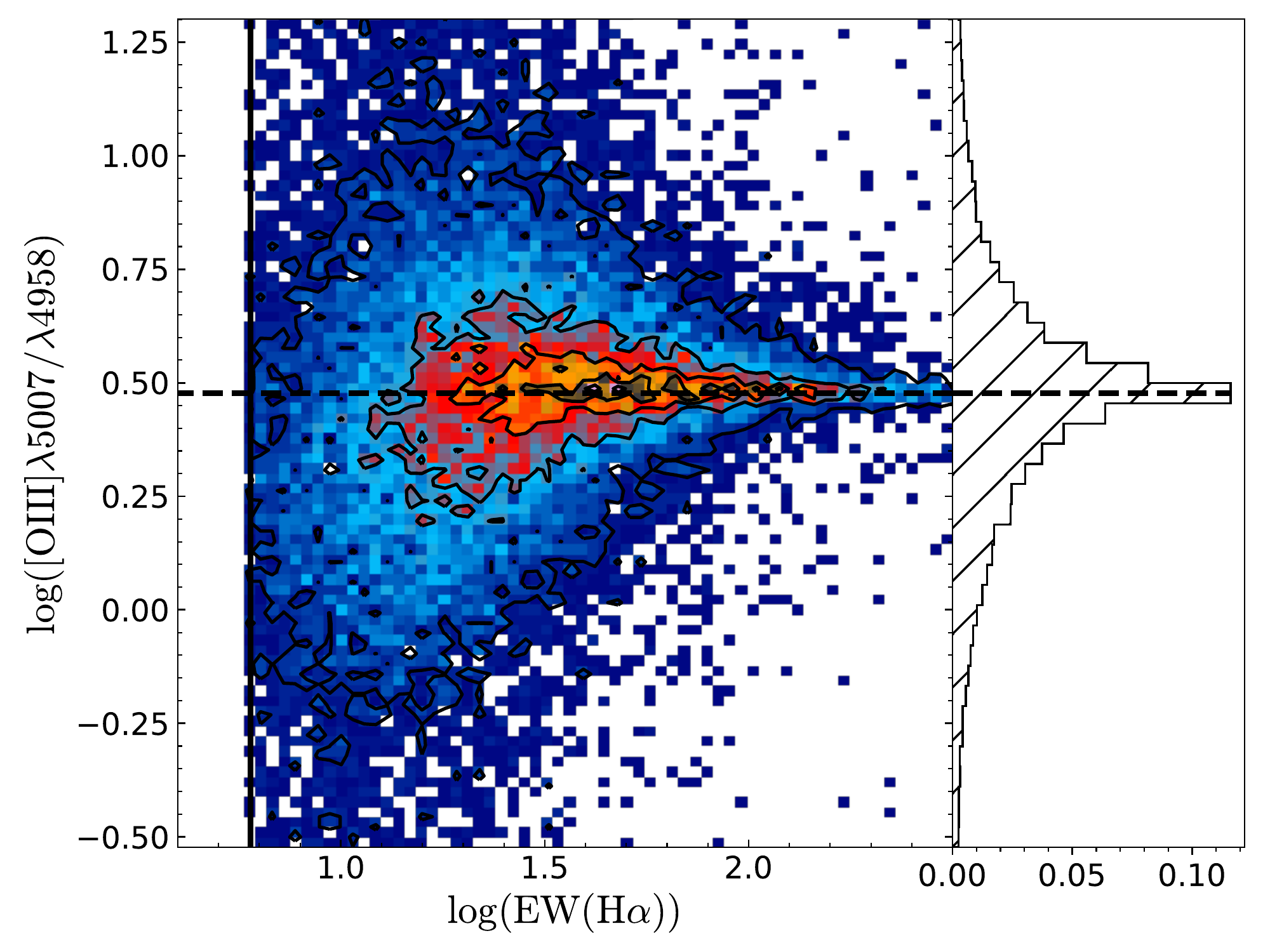}
	\hspace*{0.1cm}
	\includegraphics[width=\columnwidth]{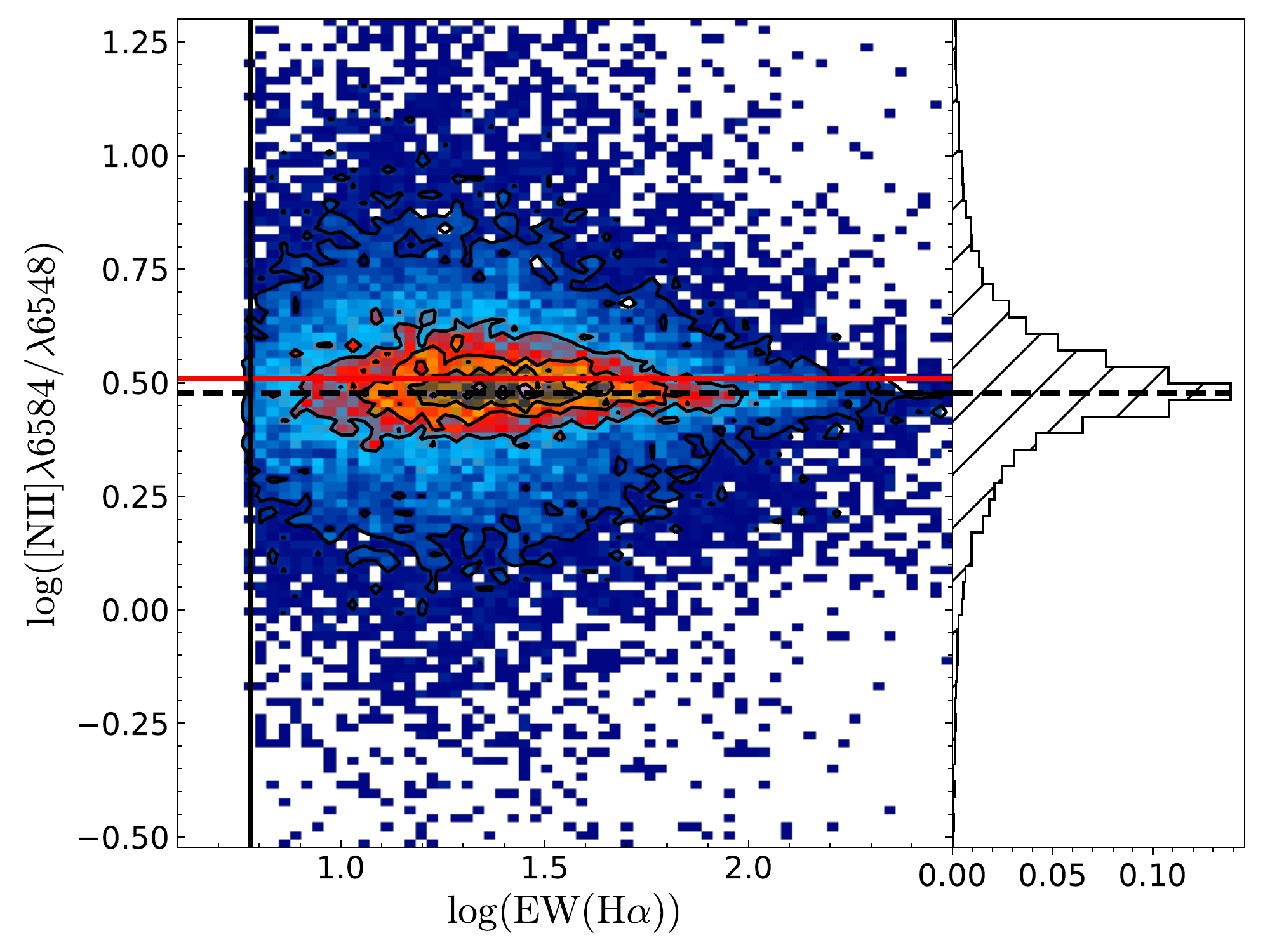}

    \caption{[OIII]$\lambda5007/\lambda4958$ and  [NII]$\lambda6584/\lambda6548$ ratios as a function of \EWHa\ for: i) \HII\ regions sample and ii) DIG sample (mean value, red diamond point) as described in the text. The red horizontal line represents the weighted-mean value used to derive the DIG correction (see \cref{sec:diff}). The black horizontal line corresponds to the theoretical value of [OIII] and [NII] lines ratio. The vertical line stands for the \EWHa threshold use to segregated the \HII\ regions from the ionized regions. On the right hand, for both panels, we show the distribution of each emission line ratios. Both distributions are centered in the expected value.
    }
    \label{fig:ratios_check}

\end{figure*}

We check the quality of our \HII\ region sample by comparing the observed distribution of the [OIII] $\lambda 5007 / \lambda4959$ and the [NII] $\lambda 6584 / \lambda6548$ flux emission lines ratios to the theoretical value \citep[$\sim 3$,][]{osterbrock89}. Therefore, these line ratios are an excellent indicator of the quality of the spectra of an ionized nebular region (and the subtraction of the underlying stellar population). \cref{fig:ratios_check} shows the density distribution of the considered line ratios as a function of the \EWHa\ for the final sample of \HII\ regions (with blue representing the low-density values). The black contours represent the density distribution; the outermost contour encloses 95\% of the regions and each consecutive one encloses a 20\% fewer number. In both panels, the horizontal line represents the theoretical value of the corresponding line ratio and the vertical line represents the \EWHa\ threshold that we used to classify the \HII\ regions (\EWHa$=6$\AA). As can be seen in the figure, most of the final selected \HII\ regions lie near the theoretical value. However, there is a significant scatter in those regions near the adopted \EWHa\ threshold. Only low S/N (i. e. low contrast at low EW) spaxels present nonphysical line ratios, as expected due to the significant errors of the measured fluxes. The actual value used to decontaminate the DIG is represented as a red horizontal line, being more compatible with expected physical value. Unfortunately, several emission lines, like [OIII]$\lambda4958$, have S$/$N <1 even when combining several DIG spaxels. Therefore, not all emission lines (and line ratios) in the spectral range of data could be corrected by DIG in our dataset.

\section{Description of the catalog}
\label{sec:cat}

The catalog of \HII\ regions described along this article, with the full set of parameters derived for each region, is presented in five different files for each galaxy (labeled by its name, GALNAME):

\begin{itemize}
    \item \texttt{seg\_Ha.GALNAME.fits}: Fits file comprising the segmentation maps indicating the actual selection of \HII\ regions. Each map has the same geometry and WCS of the original H$\alpha$ maps used to select the regions. All pixels corresponding to each \HII\ region is labeled with a single ID number that identifies that region on the catalog (for each galaxy), with IDs starting from one. Pixels labeled with zeroes correspond to areas excluded from the selection of \HII\ regions.
    \item \texttt{mask\char`_Ha.GALNAME.fits}: Fits file comprising a mask of those pixels not considered to be part of any selected \HII\ region.
    \item \texttt{HII.GALNAME.flux\_elines.csv}: CSV table including the fluxes of the emission lines extracted for each \HII\ region within the considered galaxy, together with the location of the region with respect to the center of the galaxy. Each region corresponds to one row, labeled with the ID included in the \texttt{seg\_Ha.GALNAME.fits} file and the galaxy name (GALNAME). Details of the content in this file are given in \cref{table:fe_table}.
    \item  \texttt{HII.GALNAME.SFH.csv}: CSV table including the fraction of light in the V-band of the decomposition of the stellar population currently for each stellar population included in the SSP template adopted by {\sc Pipe3D} \citep{Pipe3D_II}. Details of the content in this file are given in \cref{table:sfh_table}.
    \item \texttt{HII.GALNAME.SSP.csv}: CSV table including the average properties of the underlying stellar population (age, [Z/H], dust attenuation), as described in \citet{Pipe3D_II}. Details of the content in this file are given in \cref{table:ssp_table}.
\end{itemize}

The full catalog comprising all this files is stored in:  \url{http://ifs.astroscu.unam.mx/CALIFA/HII\_regions/} for public access. 

\begin{table*}
\centering
\begin{tabular}{lp{10cm}}
\hline
Keyword   & Description \\ \hline
\texttt{HIIREGID}   & Unique identifier of the ionized region. The generic format for all regions is \texttt{GALNAME-XX} where \texttt{GALNAME} is the name of the host galaxy and \texttt{XX} is the unique ID described in the previous section.    \\
\texttt{GALNAME}   & Name of the host galaxy        \\
\texttt{CALIFAID}  & ID assigned a each galaxy in the CALIFA survey       \\
\texttt{X}      & X centroid of the ionized region in the cube         \\
\texttt{Y} & Y centroid of the ionized region in the cube      \\
\texttt{RA} & Right ascension of the centroid of the ionized region        \\
\texttt{DEC} & Declination of the centroid of the ionized region        \\
\texttt{Ha\char`_VEL} & H$\alpha$ rotational velocity in km s$^{-1}$        \\
\texttt{Ha\char`_DISP}& H$\alpha$ rotational velocity dispersion in km s$^{-1}$         \\
\texttt{fluxOII3727} & Integrated flux of [OII]$\lambda3727$ in units of $10^{-16}$ergs s$^{-1}$cm$^{-2}$.      \\
\texttt{e\char`_fluxOII3727} & Estimated error of the integrated flux of [OII]$\lambda3727$ in units of $10^{-16}$ergs s$^{-1}$cm$^{-2}$.       \\
\texttt{EWOII3727} & Equivalent width of [OII]$\lambda3727$ in \AA       \\
\texttt{e\char`_EWOII3727} & Estimated error of equivalent width of [OII]$\lambda3727$ in \AA        \\
 .... & ...        \\
\texttt{fluxSII6731} & Integrated flux of [SII]$\lambda6731$ in units of $10^{-16}$ergs s$^{-1}$cm$^{-2}$.      \\
\texttt{e\char`_fluxSII6731} & Estimated error of the integrated flux of [SII]$\lambda6731$ in units of $10^{-16}$ergs s$^{-1}$cm$^{-2}$.       \\
\texttt{EWSII6731} & Equivalent width of [SII]$\lambda6731$ in \AA       \\
\texttt{e\char`_EWSII6731} & Estimated error of equivalent width of [SII]$\lambda6731$ in \AA        \\ \hline
\end{tabular}

\caption{The table has 213 columns, from 10th column to 213th column it comprises the flux and equivalent width data for the 51 emission lines explored in this study (irrespectively of their S/N). We have to remark that errors in the WCS are propagated in RA and DEC.}
\label{table:fe_table}

\end{table*}

\begin{table*}
\centering
\caption{Description of the \texttt{HII.GALNAME.SFH.csv} files}
\label{table:sfh_table}
\begin{tabular}{ll}
\hline
Keyword   & Description \\ \hline
COLUMN1   & HIIREGID    \\
COLUMN2   & Luminosity Fraction for age-met 0.0010-0.0037 SSP        \\
COLUMN3   & Luminosity Fraction for age-met 0.0010-0.0076 SSP        \\
...       & ...         \\
COLUMN156 & Luminosity Fraction for age-met 7.9433-0.0190 SSP      \\
COLUMN157 & Luminosity Fraction for age-met 7.9433-0.0315 SSP        \\ \hline
COLUMN158 & Luminosity Fraction for age 0.0010 SSP        \\
COLUMN159 & Luminosity Fraction for age 0.0030 SSP        \\
...       & ...         \\
COLUMN195 & Luminosity Fraction for age 12.5893 SSP      \\
COLUMN196 & Luminosity Fraction for age 14.1254 SSP       \\
COLUMN197 & Luminosity Fraction for met 0.0037 SSP        \\
COLUMN198 & Luminosity Fraction for met 0.0076 SSP        \\
COLUMN199 & Luminosity Fraction for met 0.0190 SSP        \\
COLUMN200 & Luminosity Fraction for met 0.0315 SSP        \\
COLUMN201 & Error in the Lum. Fract. for age-met 0.0010-0.0037 SSP        \\
...       & ...         \\
COLUMN399 & Error in the Lum. Fract. for met 0.0315 SSP \\ \hline
\end{tabular}
\end{table*}

\begin{table*}
\centering
\caption{Description of the \texttt{HII.GALNAME.SSP.csv} files}
\label{table:ssp_table}
\begin{tabular}{lll}
\hline
Column & Keyword   & Description \\ \hline
COLUMN1 &  \texttt{HIIREGID} &   Ionized region ID   \\
COLUMN2 &  \texttt{GALNAME}  & Galaxy name        \\
COLUMN3 &  \texttt{CALIFAID} & Name of the host galaxy        \\
COLUMN4 &  \texttt{X} & X centroid of the ionized region in the cube         \\
COLUMN5 &  \texttt{Y} & Y centroid of the ionized region in the cube      \\
COLUMN6 &  \texttt{RA} & Right ascension of the centroid of the ionized region   \\
COLUMN7 & \texttt{DEC} & Declination of the centroid of the ionized region       \\
COLUMN8 & \texttt{log\char`_age\char`_LW} & Luminosity weighted age of the stellar population in log(age/yr) \\
COLUMN9 & \texttt{log\char`_age\char`_MW} & Mass weighted age of the stellar population in log(age/yr) \\
COLUMN10 & \texttt{e\char`_log\char`_age} & Error of the age of the stellar population  \\
COLUMN11 & \texttt{ZH\char`_LW} & Luminosity weighted metallicity of the stellar population in log(Z/Z$_\odot$) \\
COLUMN12 & \texttt{ZH\char`_MW} & Mass weighted metallicity of the stellar population in log(Z/Z$_\odot$) \\
COLUMN13 & \texttt{e\char`_ZH} & Error metallicity of the stellar population  \\
COLUMN14 & \texttt{AV\char`_ssp} & Average dust attenuation of the stellar population (A$_{V,{\rm stars}}$) in mag \\
COLUMN15 & \texttt{e\char`_AV\char`_ssp} & Error of the average dust attenuation of the stellar population in mag \\
COLUMN16 & \texttt{vel\char`_ssp} & Velocity of the stellar population       in km/s \\
COLUMN17 & \texttt{e\char`_vel\char`_ssp} & Error in the velocity of the stellar population\\
COLUMN18 & \texttt{disp\char`_ssp} & Velocity dispersion of the stellar population in km/s\\
COLUMN19 & \texttt{e\char`_disp\char`_ssp} & Error in velocity dispersion of the stellar population   \\
COLUMN20 & \texttt{log\char`_ML} & Average mass-to-lightratio of the stellar population in Solar Units \\
COLUMN21 & \texttt{log\char`_Sigma\char`_Mass} & Stellar mass density in ${\rm M}_\odot/{\rm arcsec}^2$, not dust corrected     \\
COLUMN22 & \texttt{log\char`_Sigma\char`_Mass\char`_corr} & Stellar mass density in ${\rm M}_\odot/{\rm arcsec}^2$, dust corrected using the A$_{V,{\rm stars}}$ \\ \hline
\end{tabular}
\end{table*}

\section{Distribution of the \EWHa\ of the DIG}
\label{app:DIG}
\begin{figure*}
	\includegraphics[width=\textwidth]{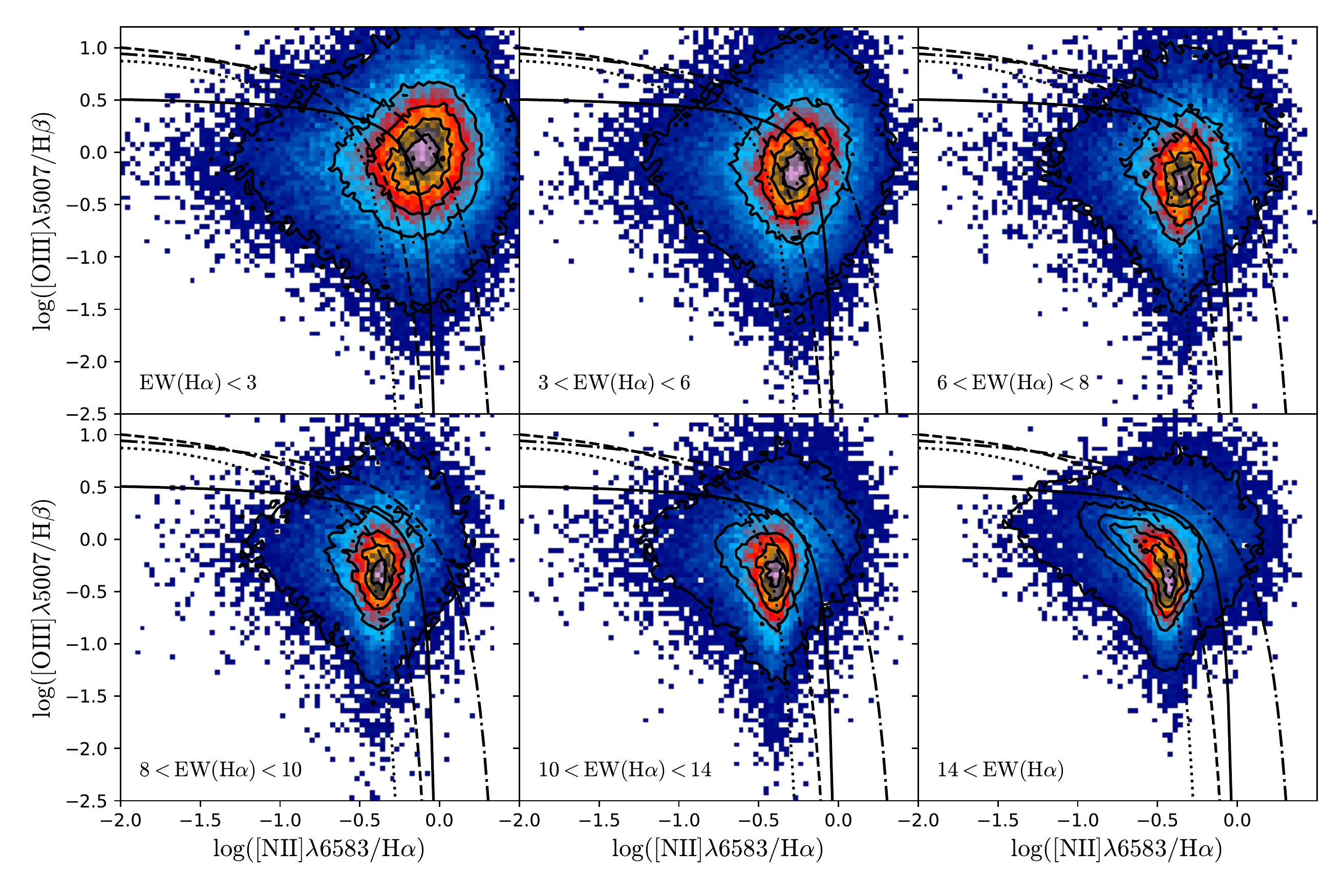}
    \caption{Final distribution of DIG spaxels segregated by mean EW(H$\alpha$) along the BPT diagram. The density contours are the same that in \cref{fig:BPT_O3N2_uncorr_corr} for the raw DIG sample. In both panels, the dotted, dot-dashed, dashed and solid lines are the demarcation curves used in \cref{fig:BPT_O3N2_uncorr_corr}.
    }
    \label{fig:BPT_O3N2_DIG_EW}
\end{figure*}

In \cref{sec:DIG_char}, the properties of DIG spaxels from all galaxies were discussed. In order to enhance the DIG's characterization, we apply selection criteria to galaxies, as we discussed above. \cref{fig:BPT_O3N2_DIG_EW} shows the distribution of \textit{clean} DIG sample on the BPT diagram segregated by the EW(H$\alpha$). The code color represents the points density and the black contours represent the distribution of raw DIG sample.  As can be seen in the figure, for low EWs values, the spaxels distribution is more located towards the area associated with ionization by the old stellar population. 
On the other hand, higher EWs values, the distribution is located in the SF area of the diagram. Along the intermediate range, the spaxel distribution moves from the area associated with ionization by the old stellar populations towards the SF area. This trend is also observed in the distribution of the raw DIG sample (black contours). However, a significant fraction of spaxels lies below the Kauffman's curve for raw DIG sample (black contours) and clean DIG sample. They correspond to spaxels contaminated by ionization by \HII\ regions due to PSF wings. Although, the contribution of these types of spaxels was reduced considerably in the \textit{clean} DIG sample. Unfortunately, the spaxels in the SF zones  are still present across all EW values. Hence, we cannot segregate the diffuse gas ionized by leaking of ionizing photons from the \HII\ regions due to the PSF wings contribution.

%%%%%%%%%%%%%%%%%%%%%%%%%%%%%%%%%%%%%%%%%%%%%%%%%%

% Don't change these lines
\bsp	% typesetting comment
\label{lastpage}
\end{document}